\documentclass[a4paper,useAMS,usenatbib, onecolumn,fleqn]{mn2e}

\usepackage{graphics}
\usepackage[dvipdfm]{graphicx}

\usepackage{amsmath}
\usepackage{amssymb}
\usepackage{bm}
\usepackage{color}
\usepackage{tabularx}
\usepackage{epstopdf}
\usepackage{fancyvrb}
\usepackage{longtable}
\usepackage{arydshln}

\def\ltsima{$\; \buildrel < \over \sim \;$}
\def\lsim{\lower.5ex\hbox{\ltsima}}
\def\gtsima{$\; \buildrel > \over \sim \;$}
\def\gsim{\lower.5ex\hbox{\gtsima}}
\def\ga{\mathrel{\hbox{\rlap{\hbox{\lower4pt\hbox{$\sim$}}}\hbox{$>$}}}}
\def\la{\mathrel{\hbox{\rlap{\hbox{\lower4pt\hbox{$\sim$}}}\hbox{$<$}}}}

\def\be{\begin{equation}}
\def\ee{\end{equation}}
\def\ba{\begin{eqnarray}}
\def\ea{\end{eqnarray}}
\def\bv{\begin{verbatim}}
\def\ev{\end{verbatim}}

\def\erg{\rm erg}

\def\sec{\rm sec}
\def\cm{\rm cm}

\def\Tb{T_{\rm b}}
\def\Ts{T_{\rm s}}
\def\Tk{T_{\rm k}}
\def\Tg{T_{\rm g}}
\def\Ta{T_{\alpha}}
\def\Tcmb{T_{\rm CMB}}

\def\yc{y_{\rm c}}
\def\ya{y_{\alpha}}

\def\kB{k_{\rm B}}
\def\hp{h_{\rm p}}
\def\A10{A_{10}}
\def\v10{\nu_{10}}
\def\k10{\kappa_{10}}
\def\dTb{\delta T_{\rm b}}

\def\HI{H\;{\sc i}}
\def\HII{H\;{\sc ii}}

\def\HeI{He\;{\sc i}}
\def\HeII{He\;{\sc ii}}
\def\HeIII{He\;{\sc iii}}
\newcommand{\iHeII}{$^3$He\;{\sc ii}}

\def\He3{^{3}{\rm He}}
\def\iHe3{^{3}{\rm He}^{+}}

\def\nel{n_{e}}
\def\nH{n_{\rm H}}
\def\nHII{n_{\rm H{\small II}}}
\def\nHI{n_{\rm H{\sc I}}}
\def\nHII{n_{\rm H{\sc II}}}
\def\nHe{n_{\rm He}}
\def\nHeI{n_{\rm He{\sc I}}}
\def\nHeII{n_{\rm He{\sc II}}}
\def\nHeIII{n_{\rm He{\sc III}}}

\title[Probing large scale filaments with H{\sc ~i} and $^3{\rm
    He}${\sc ~ii}]{Probing large scale filaments with {\bf H}{\sc ~i}
  and $^3{\bf He}${\sc ~ii}}
\author[]{
Yoshitaka Takeuchi$^{1}$\thanks{E-mail:yoshitaka@nagoya-u.jp}, 
Saleem Zaroubi$^{2}$ and 
Naoshi Sugiyama$^{1,3,4}$
\vspace{5pt} \\
$^{1}$Department of physics, Nagoya University, 
Naogya 464-8602, Japan \\
$^{2}$Kapteyn Astronomical Institute, University of Groningen, 
P.O. Box 800, 9700 AV Groningen, The Netherlands \\
$^{3}$Kobayashi-Maskawa Institute, Nagoya University, Nagoya 464-8602,
Japan \\
$^{4}$Kavli Institute for the Physics and Mathematics of the Universe,
University of Tokyo, Kashiwa 277-8568, Japan 
}

\begin{document}

\date{Accepted, Received; in original form }

\pagerange{\pageref{firstpage}--\pageref{lastpage}} \pubyear{2014}

\maketitle

\label{firstpage}

\begin{abstract}

We explore the observability of the neutral hydrogen (\HI) and the
singly-ionized isotope helium-3 ($^3$\HeII) in the intergalactic
medium (IGM) from the Epoch of Reionization down to the local
Universe. The hyperfine transition of $^3$\HeII, which is not as well
known as the \HI\ transition, has energy splitting corresponding to
8~cm. It also has a larger spontaneous decay rate than that of neutral
hydrogen, whereas its primordial abundance is much smaller.  Although
both species are mostly ionized in the IGM, the balance between
ionization and recombination in moderately high density regions
renders them abundant enough to be observed.  We estimate the emission
signal of both hyperfine transitions from large scale filamentary
structures and discuss the prospects for observing them with current
and future radio telescopes. We conclude that \HI\ in filaments is
possibly observable even with current telescopes after 100 hours of
observation. On the other hand, \iHeII\ is only detectable with future
telescopes, such as SKA, after the same amount of time.

\end{abstract}


\section{Introduction}
\label{sec:intro}

Since the prediction of the 21~cm hyperfine transition by
\citet{Hulst:1945} and its first detection by \citet{Ewen:1951} and
\citet{Muller:1951}, forbidden quantum transition lines have been
powerful tools in exploring various astrophysical systems. 
The advent of new larger and more sensitive radio telescopes makes it
possible to use such tools for exploring even higher redshifts and
lower density environments.  Two recent examples are the use of
redshifted 21 cm for exploring the Epoch of Reionization (EoR) (see
e.g., \citet{Furlanetto:2006,Pritchard:2012, Zaroubi:2013}); and for
mapping neutral gas around redshift $\sim 1$-$2$ to probe the baryon
acoustic peaks \citep{Chang:2008}. 
Another line, that has been discussed in the literature, is the
singly-ionized helium-3 isotope, $^3$\HeII\ \citep{Townes:1957,
  Sunyaev:1966,Goldwire:1967,Rood:1979,Bell:2000,
  McQuinn:2009b,Bagla:2009}.  In this study we propose to use \HI\ and
\iHeII\ to detect large scale filaments and the accumulation of
baryons within them as a function of redshift.

At the local Universe, most of the baryons in the intergalactic space
reside in filamentary structures, the so-called cosmic web
\citep{Bond:1996}. These filaments are readily seen by modern galaxy
surveys, such as the 2-degree Field Galaxy Redshift Survey
\citep{Colless:2001, Erdogdu:2004}, the Sloan Digital Sky Survey
\citep{York:2000} and the 2-Micron All-Sky Survey
\citep{Skrutskie:2006}.
Unfortunately, however, such surveys are not able to probe the
baryonic content of these filaments and its distribution, where less
than half of baryons at the local Universe have been identified
\citep{Cen:1999, Fukugita:2004}.

The observation of filamentary structures in the IGM through the
hyperfine transition of neutral hydrogen can be a powerful tool that
has the potential for detecting the missing baryons in the local
Universe (see e.g., \citet{Popping:2009} and \citet{Popping:2011}).
However, below redshift $\sim 6$, the detection of \HI\ in the diffuse
IGM becomes very difficult as the Universe reionizes and a very small
neutral fraction is left. Despite this difficulty, \citet{Chang:2010}
and \citet{Masui:2012} have detected \HI\ at low redshift by cross
correlating the aggregate 21-cm glow with data from other probes of
large-scale structure. Still, the signal of the auto-correlation at
low-redshifts can be interpreted as an upper bound on the 21 cm signal
\citep{Switzer:2013}.
Direct detection of \HI\ emission signal from IGM filaments has not
been reported yet, but Ly-$\alpha$ forest absorption towards
background quasars implies \HI\ from filaments.

Observation of \iHeII\ transition from the diffuse and filamentary
structure in the IGM is very difficult to carry out, mainly because of
its low abundance relative to hydrogen, and therefore, in principle in
harder to detect. Furthermore, the sensitivity of the current radio
telescopes at the appropriate frequency range is too poor for such a
task.
However, despite the low primordial abundance of $^3{\rm He}$
relative to hydrogen ($\sim 10^{-5}$), it has some mitigating factors
that render it observationally accessible. These factors are as
follows:
\begin{enumerate}
\item The spontaneous decay rate of $^3$\HeII\ ($\A10 = 1.959 \times
  10^{-12}~{\rm s}^{-1}$) is $\sim$680 times larger than that of the
  \HI\ ($\A10 = 2.876 \times 10^{-15}~{\rm s}^{-1}$)
  \citep{Gould:1994,Sunyaev:1966,Goldwire:1967}, which significantly
  increases its emission rate.
\item The ionization energy of \HeII\ ($54.4$~eV) is 4 times larger
  than that of neutral hydrogen ($13.6$~eV), namely, it requires
  harder photons to fully ionize. Conversely, this difference in
  ionization energy causes the \HeIII\ recombination rate to be
  $\sim$5 times larger than that of \HII\ \citep{Verner:1996}.  Hence,
  the abundance fraction of \iHeII\ in the IGM should be larger than
  that of \HI.
\item The line transition occurs at a frequency of $8.7$~GHz, in which
  the foreground synchrotron radiation from our galaxy and distortions
  from the terrestrial ionosphere are both less pronounced.
\end{enumerate}

The overdensity of filaments in the IGM is roughly of the order of
10-100 times the mean density of the Universe. Therefore, given the
size of such filaments, one can accumulate a sizable column density of
both \HI\ and \iHeII, in particular if they are elongated along the
line of sight.
Furthermore, at these densities, the recombination rates for both
species are generally shorter than Hubble time and a reasonable
fraction of \HI\ and \HeII\ is expected to be present. Detection of
either species will go a long way in accounting for the baryons in the
Universe at the redshift of detection.
In addition, the relative abundance of these species can constrain the
hardness of the UV background as a function of redshift.

Our aim in this study is to estimate the prospect of observing these
two species within large scale structure filaments with the present
and future radio telescopes in the redshift range $\sim 0-8$. Here, we
consider a number of single dish telescopes, e.g.,
GBT\footnote{https://science.nrao.edu/facilities/gbt}
\citep{Chang:2010,Masui:2012} and
Arecibo\footnote{http://www.naic.edu/}\citep{Freudling:2011}, and
radio interferometers such as
EVLA\footnote{https://science.nrao.edu/facilities/vla}, and 
GMRT\footnote{http://gmrt.ncra.tifr.res.in/}. 
Note that LOFAR does not have the proper frequency range for observing either 
of the two transitions, except \HI\ at $z\gsim 6$ \citep{Haarlem:2013}. 
We also make predictions for future telescopes that generally have
more sensitivity, a larger field of view and wider frequency coverage:
FAST\footnote{http://fast.bao.ac.cn/en/FAST.html}, 
MeerKAT\footnote{http://www.ska.ac.za/meerkat/index.php} and, 
of course, the mega radio telescope SKA\footnote{http://www.skatelescope.org/}, 
which is expected to be completed around 2024.

This paper is organized as follow. 
We summarize the physical models used in the paper for calculating the
\HI\ and \iHeII\ signal from large scale filamentary structure in the
IGM in Section~\ref{sec:model}. In the same section, we also evaluate
the abundance of each ionization state, the gas temperature in the IGM
and the spin and brightness temperatures.
The predicted hyperfine transition signal of \HI\ and $^3$\HeII\ for
simple models as well as for filaments from large scale structure
simulations is presented in Section~\ref{sec:signal}.
Finally we discuss the prospects of observing the emission signal of
the hyperfine transition from filamentary structures with current and
future radio telescopes in Section~\ref{sec:obs}. The paper ends with
a conclusion and discussion section.

Throughout this paper, we adopt the cosmological parameters from the
WMAP 7 years  data \citep{Komatsu:2011}.

\section{The physical Model for \HI\ and $^3$\HeII }
\label{sec:model}

To calculate the signal of \HI\ and $^3$\HeII , an evaluation of their
ionization states and abundances is required. To achieve this, we
solve the radiation balance equations between the ionization and
recombination processes. These processes are highly dependent on the
environment in the IGM, e.g. the density and the temperature of gas.
The photo-ionization and the heating processes through the IGM are
also determined by the background radiation field. The UV/X-ray
background model adopted in this paper is the 
\citet[hereafter HM12]{Haardt:2012} model.

The observed signal of the hyperfine transition is related to the
relative occupation number of the excited state relative to the ground
state, which determines the spin temperature.  Other than the CMB, two
processes affect the relative occupation number density of hyperfine
state: the collisional excitation process and the Ly-$\alpha$ pumping
of the line, also known as the Wouthuysen-Field process
\citep{Wouthuysen:1952,Field:1958,Field:1959b}.  We estimate the spin
temperature taking into account both these processes.

In this section, the physical equations needed for the estimation of
the hyperfine transition are summarize and the signal expected from a
simple slab model that represents the filamentary structure in the IGM
is calculated.  Specifically, in Subsec.~~\ref{sec:ion} and
\ref{sec:gas},
we calculate the evolution of the abundance for each ionization state
and the gas temperature in the IGM following the equations in
\citet{Fukugita:1994}, where we adopt the photo-ionization rate
$\Gamma(z)$ and the heating function ${\cal H}(z)$ given by HM12.  
In Subsec.~\ref{sec:spin}, the physical processes determining the spin
state and estimate the spin temperature are summarized.
Finally, the brightness temperature expected from a simple filamentary
structure model is shown in subsec.~\ref{sec:bright}.

\subsection{Ionization state of hydrogen and helium}
\label{sec:ion}

We here assume that the system contains only H and He. The abundances
of each H and He species, i.e., \HI , \HII , \HeI , \HeII\ and \HeIII
, are given by solving the balance equation between the ionization and
recombination processes \citep{Fukugita:1994}. The equation for the
hydrogen species is:
\begin{equation}
	\frac{d}{dt} \left[ \frac{n_{\rm HII}}{\nH} \right]
        = \Gamma_{\rm H{\small I}} \nel \frac{\nHI}{\nH} 
        + \beta_{\rm H{\small I}} \nel \frac{\nHI}{\nH}
        - \alpha_{\rm H{\small II}} \nel \frac{\nHII}{\nH} \, ,
\label{eq:ionHII}
\end{equation}
where $n_e$ is the number density of electron, $\nHI$ and $\nHII$ are
respectively the number densities of the neutral and the ionized
hydrogen, and $\nH \equiv \nHI + \nHII$ is the total number density of
hydrogen.

The balance equations for \HeI , \HeII\ and \HeIII\ are:
\begin{equation}
	\frac{d}{dt} \left[ \frac{n_{\rm HeII}}{\nHe} \right]
        = \Gamma_{\rm He{\small I}} \nel \frac{\nHeI}{\nHe} 
        + \beta_{\rm He{\small I}} \nel \frac{\nHeI}{\nHe} 
        - \beta_{\rm He{\small II}} \nel \frac{\nHeII}{\nHe} 
        - (\alpha_{\rm He{\small II}} + \xi_{\rm He{\small II}}) \nel \frac{\nHeII}{\nHe} 
        + \alpha_{\rm He{\small III}} \nel \frac{\nHeIII}{\nHe} \, ,  
\label{eq:ionHeII}
\end{equation}
and
\begin{equation}
	\frac{d}{dt} \left[ \frac{n_{\rm HeIII}}{\nHe} \right]
        = \Gamma_{\rm He{\small II}} \nel \frac{\nHeII}{\nHe} 
        + \beta_{\rm He{\small II}} \nel \frac{\nHeII}{\nHe} 
        - \alpha_{\rm He{\small III}} \nel \frac{\nHeIII}{\nHe} \, ,
\label{eq:ionHeIII}
\end{equation}
where $\nHeI$, $\nHeII$ and $\nHeIII$ are respectively the number
densities of the neutral, the singly-ionized and the doubly-ionized
helium and  $\nHe \equiv \nHeI + \nHeII + \nHeIII$ is the
total number density of helium.  
Assuming ionization equilibrium, the electron number density is given by 
\begin{equation}
  n_e = \nHII + \nHeII + 2 \nHeIII \, .
\label{eq:electron}
\end{equation}

The functions used in the ionization equations are the
photo-ionization rate $\Gamma_{\rm X}$, the recombination rate
$\alpha_{\rm X}$, the dielectronic recombination rate $\xi_{\rm X}$,
and the collisional ionization rate $\beta_{\rm X}$.
The label ${\rm X}$ stands for the species used in this calculation,
${\rm X} \in\ $\{\HI, \HII, \HeI, \HeII, \HeIII\}. We adopt the value
of each photo-ionization state $\Gamma_{\rm X}$ from HM12, which takes
into account the photo-ionization heating of \HI , \HeI\ and \HeII\ ,
and Compton heating.
The other functions adopted in our calculation, $\alpha_{\rm X}$,
$\beta_{\rm X}$ and $\xi_{\rm X}$, and are summarized in
Appendix~\ref{sec:rates} \citep{Spitzer:1978,Verner:1996}.  These
values depend on the gas temperature and one has to compute the
evolution of the gas temperature simultaneously together with the set
of above four equations for the above mentioned ionization states.

\begin{figure}
\begin{center}
\includegraphics[clip,keepaspectratio=true,width=0.75
  \textwidth]{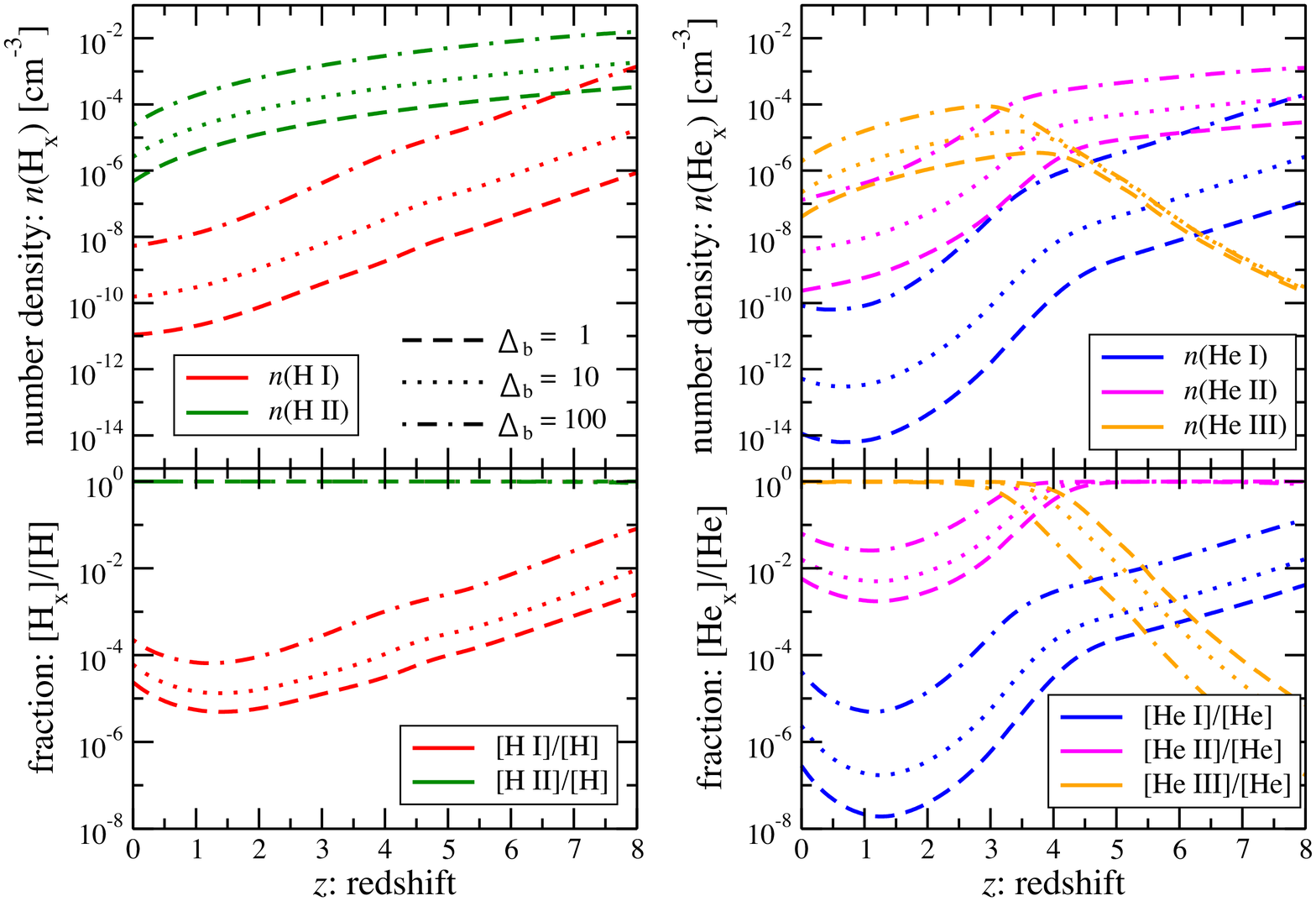}
\caption[]{The redshift evolution of the ionization state of hydrogen
  (Left) and helium (Right). The top panels show the number density of
  each ionization state, while the bottom panels show their fractional
  abundance against the total number of hydrogen $n_{{\rm H_x}}/n_{\rm
    H}$ or helium $n_{\rm He_x}/n_{\rm He}$; the label ${\rm X}$
  stands for each atomic ionization state. The lines corresponding to
  each species are as follows: \HI\ (red), \HII\ (green) \HeI\ (blue),
  \HeII\ (magenta) and \HeIII\ (orange).  The different line types
  represent the different values of the baryon density contrast
  parameter $\Delta_{\rm b}$ defined in Eq.\,(\ref{eq:deltab}), and
  dashed, dotted and dashed-dot lines, respectively, correspond to
  $\Delta_{\rm b}=$1, 10 and 100.  }
\label{fig:ion}
\end{center}
\end{figure}

\subsection{Gas temperature}
\label{sec:gas}

While the evolution of each ionization state depends on the
temperature of gas in the IGM, the gas temperature also depends on the
value of the local density and the flux of the background radiation
through the heating or cooling processes.
Assuming that the IGM is in the thermal equilibrium, which means the
electrons, ions and neutrals have a common temperature, the gas
temperature can be characterized by the electron temperature and the
entropy equation can be written as \citep{MoWhite:2010}:
\begin{equation}
  \frac{d \ln \Tg}{d \ln (1+z)}  
  = (\gamma - 1) \left[ 3 + \frac{1}{(\gamma -1)} \frac{d \ln \mu}{d\ln (1+z)}
    - \frac{{\cal H} - \Lambda}{H(z) \, n \, \kB \Tg} \right]  \, , 
\label{eq:Tg}
\end{equation}
where $n$ is the baryon number density, $\gamma$ is the adiabatic
index, $\mu$ is the mean molecular weight, and $H(z)$ is the Hubble
parameter. ${\cal H}$ and $\Lambda$ are the heating and cooling
functions, respectively. 
We adopt the HM12 heating function ${\cal H}$, wheres the cooling
function used in this work is summarized in Appendix~\ref{sec:cooling}
\citep{Black:1981a,Cen:1992c}.
The first term in the right hand side of Eq.~(\ref{eq:Tg}) corresponds
to the adiabatic cooling through the Hubble expansion.

To describe the local density contrast of baryons, we define the local
number density of each baryon component as,
\begin{equation}
  n_{\rm X} 
  \equiv (1+\delta) \, f_{\rm b} \, \bar{n}_{\rm X} 
  = \Delta_{\rm b} \, \bar{n}_{\rm X} \, ,
\label{eq:deltab}
\end{equation}
where $\delta$ is the matter density fluctuations, $f_{\rm b}$
represents the bias, $\bar{n}_{\rm X}$ is the average number density
of a component ${\rm X}$, and the baryon density contrast is
$\Delta_{\rm b}$ (defined as $(1+\delta)\,f_{\rm b}$). Although the
distribution of baryons might be different from that of matter, this
effect is assumed to be small on the scales of interest in this work
i.e. $f_{\rm b}=1$.

The interaction time scales of the ionization and recombination for
each species are given by 
$t^{\rm ion}_{\rm X}=(\Gamma_{\rm X} + \beta_{\rm X} n_e)^{-1}$ and 
$t^{\rm rec}_{\rm X}=(\alpha_{\rm X} n_e)^{-1}$, respectively.
If these time scales are much shorter than Hubble time, the assumption
of the ionization equilibrium is valid. For example, the typical time
scales for hydrogen in the IGM are 
$t^{\rm ion}_{\rm HI}\sim (10^{-14}/\Gamma_{\rm HI}) \times 10^7 $~years and 
$t^{\rm rec}_{\rm HII}\sim \Delta_{\rm b}^{-1} \times 10^{11}$~years at $z=0$ for
ionization and recombination, respectively. Since the recombination
time scale is longer than the Hubble expansion time within low-density
regions, the assumption of the ionization equilibrium is not satisfied
for hydrogen.
However, if the photoionization rate is much larger than the
recombination rate, the timescale is actually quite short compared to
the Hubble time and the assumption of equilibrium is reasonably
satisfied.
Furthermore, the assumption of ionization equilibrium is valid for
high-densities ($\Delta_{\rm b} \ga 10$), such as in filamentary
structures in the IGM.  In such a region, the local thermal
equilibrium is also reasonably satisfied, since the interactions
between the electrons and ionized species work effectively.

Assuming equilibrium, we solve these five independent equations
Eqs.\,(\ref{eq:ionHII})-(\ref{eq:Tg}) simultaneously.
The results of the abundance for the baryon density parameter,
$\Delta_{\rm b} =$ 1, 10 and 100 are shown in
Figure\,\ref{fig:ion}. Note that the assumption of the equilibrium is
invalid for $\Delta_{\rm b}=1$, but we still show it for comparison
purposes.

\subsection{Spin temperature}
\label{sec:spin}

In this subsection, we discuss the evaluation of the spin temperature
and its coupling factors due to collisions and to the Wouthuysen-Field
effect for both neutral hydrogen \HI\ and singly-ionized isotope
helium-3 $^3$\HeII.

The relative abundance of the hyperfine excited state and the ground
state is determined by the spin temperature $\Ts$ as 
\begin{equation}
  \frac{n_1}{n_0} =
  (g_1/g_0)\exp\left[ - \frac{\hp\nu_{10}}{\kB \Ts} \right] \, ,
\end{equation}
where $n_1$ and $n_0$ are the numbers of hyperfine excited state and
the ground state, respectively. The factor $g_1/g_0$ is the
statistical weight of first-excited/ground state, 
$\hp$ is the Planck constant,
$\kB$ is the Boltzmann constant,
$\nu_{10}$ is the frequency which corresponds to the energy of the
hyperfine transition.

The spin temperature is determined by three processes. The first is
coupling to the CMB temperature which, due to to the Rayleigh-Jeans
tail, can excite the hyperfine line. The second is the collisional
excitation or de-excitation of the spin states. The third is the
change of the spin states through the absorption and spontaneous
re-emission of a Ly$\alpha$ photon or any Lyman-series photon
\citep{Madau:1997,Shaver:1999,Hirata:2006,Furlanetto:2006a,Pritchard:2006}.
The first process couples the spin temperature to the CMB photons,
whereas the other two processes couple it to the gas. As a result the
spin temperature is \citep{Field:1958}
\begin{equation}
	\Ts = \frac{(\Tcmb + \yc \Tk + \ya \Ta)}{(1+\yc + \ya) } \, ,
\label{eq:tspin}
\end{equation}
where $\Tcmb$ is the CMB temperature,
$\Tk$ is the kinetic temperate of gas, 
$\Ta$ is the brightness temperature of the Ly$\alpha$ radiation field, 
$\yc$ and $\ya$ are the coupling factors of the collisional process and the Wouthuysen-Field
process. The coupling factors can be
written as 
\begin{equation}
	\yc = \frac{C_{10}}{A_{10}} \frac{T_{*}}{\Tk} \, , 
        \qquad
	\ya = \frac{P_{10}}{A_{10}} \frac{T_{*}}{\Ta} \, , 
\end{equation}
where 
$A_{10}$ is the spontaneous decay rate from state 1 to 0,
$C_{10}$ is the rate of collisional de-excitation, $P_{10}$ is the
rate of de-excitation due to the absorption of a Ly$\alpha$ photon, and  
$T_{*}$ is the equivalent temperature defined as 
$T_{*} \equiv \hp \nu_{10}/\kB$.

In general, the rate of collisional de-excitation is written as 
\begin{equation}
  C_{10} = n_e \sqrt{\frac{\kB \Tk}{\pi m_e c^2}} c \bar{\sigma} \, ,
\end{equation}
and $\bar{\sigma}$ is the averaged cross-section for spin exchange, given by 
\begin{equation}
  \bar{\sigma} = \frac{1}{(\kB \Tk)^2} \int_0^{\infty} dE 
  \sigma(E) E e^{-E/(\kB \Tk)} \, ,
\end{equation}
where $\sigma(E)$ is the cross-section for spin exchange as a function
of collision energy.

The rate of de-excitation due to the
absorption of Ly$\alpha$ photons is 
\begin{equation}
	P_{10} = \frac{4 \pi e^2 f_\alpha}{m_{e} c} 
        \left\{
        \begin{array}{ll}
          \dfrac{4}{27} J_\alpha & \mbox{for \HI}  \\ [8pt]
          \dfrac{4}{9} J_\alpha  & \mbox{for $^3$\HeII}
        \end{array}
        \right. 
\label{eq:P10}
\end{equation}
where $e$ is the electron charge, $m_e$ is the electron mass,
$f_\alpha$ is the oscillator length of the Ly$\alpha$ transition,
$J_\alpha$ is the flux at Ly$\alpha$ wave-length, and $g_1/g_0$ is the
statistical weight of the first-excited/ground state.
These values are defined for each hyperfine structure and we summarize
some values related to these processes in Table~\ref{tb:hyperfine}. We
give the detailed values on these rates for \HI\ and $^3$\HeII\ in the
following sections.


\renewcommand{\thefootnote}{\fnsymbol{footnote}}

\begin{table}
\begin{center}
\begin{tabular}{cccccccc}
\hline \hline
 Species 
& \makebox[15mm][c]{$\lambda_{10}$ [cm]} 
& \makebox[15mm][c]{$\nu_{10}$ [MHz]} 
& \makebox[15mm][c]{$A_{10}$ [s$^{-1}$]}
& \makebox[15mm][c]{$\lambda_\alpha$ [${\rm \AA}$]}
& \makebox[15mm][c]{$f_\alpha$}
& \makebox[15mm][c]{$g_1/g_0$}
& \makebox[35mm][c]{primordial abundance} 
\\
\hline 
\HI & 
21.1 & 1420.4 & $2.876 \times 10^{-15}$ 
& 1215.67
& 0.4162
& 3/1
& --- 
\\
$^3$\HeII & 
3.46 & 8665.7 & $1.959 \times 10^{-12}$ 
& 303.78
& 0.4162
& 1/3
& $1.0 \times 10^{-5}$ 
\\
\hline \hline
\end{tabular}
\end{center}
\caption[]{The parameters related to the hyperfine transitions for
  \HI\ and $^3$\HeII. $\lambda_{10}$ and $\nu_{10}$ are the
  wave-length and frequency corresponding to the energy of the
  hyperfine transition, $A_{10}$ is the spontaneous decay
  rate\footnotemark[2], $\lambda_\alpha$ and $f_\alpha$ are the
  wavelength and the oscillator length of the Ly$\alpha$ transition,
  respectively, and $g_1/g_0$ is the statistical weight of
  first-excited/ground state. The primordial abundance for $^3{\rm
    He}$ is expressed as the fractional abundance against hydrogen,
  i.e. $^3{\rm He}/{\rm H}$. }
\label{tb:hyperfine}
\end{table}

\footnotetext[2]{The values of $A_{10}$ for some of the materials which is
  interesting on radio astronomy can be found in
  \citep{Townes:1957,Gould:1994,Sunyaev:1966,Goldwire:1967}.}
\setcounter{footnote}{0}
\renewcommand{\thefootnote}{\arabic{footnote}}

\subsubsection{\HI\ : neutral hydrogen}

For collisional excitations of neutral hydrogen (\HI), the main
process is the collision with electrons, while collisions with protons
or other neutral hydrogen atoms are sub-dominant.  The total rate of
collisional de-excitation can be expressed as the summation of these
three processes;
\begin{equation}
	C_{10}^{{\rm H\;{\sc I}}} = \left[ \kappa_{10}^{{\rm HH}}(\Tk) n_{\rm H} 
        + \kappa_{10}^{e{\rm H}}(\Tk) n_{e} + \kappa_{10}^{p{\rm H}}(\Tk) n_{p} \right] \, ,
\label{eq:C10_HI}
\end{equation}
where $\kappa_{10}^{{\rm HH}}$, $\kappa_{10}^{e{\rm H}}$ and
$\kappa_{10}^{p{\rm H}}$ are the collisional rates of 
${\rm H}$-${\rm H}$, $e$-${\rm H}$ and $p$-${\rm H}$ processes, respectively.
One can can find the collisional rates as functions of the kinetic
temperature $\Tk$ in
\citet{Zygelman:2005,Sigurdson:2006,Furlanetto:2007a,Furlanetto:2007b}.

For the Wouthuysen-Field process of \HI, following Eq.\,(\ref{eq:P10}), the
de-excitation rate is related to the radiation field as 
\begin{equation}
	P_{10}^{\rm H{\small I}} 
        = \frac{16 \pi e^2 f_\alpha^{\rm H{\small I}}}{27 m_{e} c} 
        J_{{\rm Ly}\alpha,{\rm H{\small I}}} \, ,
\label{eq:P10_HI}
\end{equation}
where $f_\alpha^{\rm H{\small I}}=0.4162$ is the oscillator length of
the \HI\ Ly$\alpha$ transition, and $J_{{\rm Ly}\alpha,{\rm H{\small I}}}$
is the flux at \HI\ Ly$\alpha$ wave-length 
($\lambda_{{\rm Ly}\alpha,{\rm H{\small I}}}=1216 {\rm \AA}$).

\subsubsection{$^{3}$\HeII\ : singly-ionized isotope helium-3}

The most dominant process of collisional coupling for the
singly-ionized isotope helium-3 ($^3$\HeII) is collisions with
electrons.  
Then, following Eq.\,(\ref{eq:C10_HI}), the collisional
rate can be written as
\begin{equation}
  C_{10}^{^3{\rm He{\small II}}} 
  = n_e \sqrt{\frac{\kB \Tk}{\pi m_e c^2}} c \, \bar{\sigma}^{e{^3{\rm He}}} \, ,
\label{eq:C10_3HeII}
\end{equation}
where $\bar{\sigma}^{e\,{^3{\rm He}}}$ is the average cross-section of spin
exchange between $^3$\HeII\ and electrons, which can be approximated
as  \citep{McQuinn:2009b},
\begin{equation}
  \bar{\sigma}^{e\,{{^3\rm He}}} 
   \simeq \frac{14.3 {\rm eV}}{ \kB \Tk} a_o^2 \, ,
\end{equation}
where $a_o$ is the Bohr radius. 

For the Wouthuysen-Field process of $^3$\HeII, the de-excitation rate can be
estimated from Eq.\,(\ref{eq:P10}) in the same manner as the case of
\HI\ and given by
\begin{equation}
	P_{10}^{^3{\rm He{\small II}}} 
        = \frac{16 \pi e^2 f_\alpha^{\rm He{\small II}}}{9 m_{e} c}
        J_{{\rm Ly}\alpha,{\rm He{\small II}}} \, ,
\label{eq:P10_3HeII}
\end{equation}
where 
$f_\alpha^{\rm He{\small II}} = 0.4162$ is the oscillator length of
the \HeII\ Ly$\alpha$ transition, and $J_{{\rm HeII},{\rm Ly}\alpha}$
is the flux at \HeII\ Ly$\alpha$ wavelength 
($\lambda_{{\rm Ly}\alpha,{{\rm He\small II}}}=304S {\rm \AA}$). 
In both cases, i.e. for \HI\ and $^3$\HeII, one has to assume a model
for the Ly-$\alpha$ flux as a function of redshift.  Throughout this
paper, the recent model of UV/X-ray background from HM12 is used.

\begin{figure}
\begin{center}
\includegraphics[clip,keepaspectratio=true,width=0.45
  \textwidth]{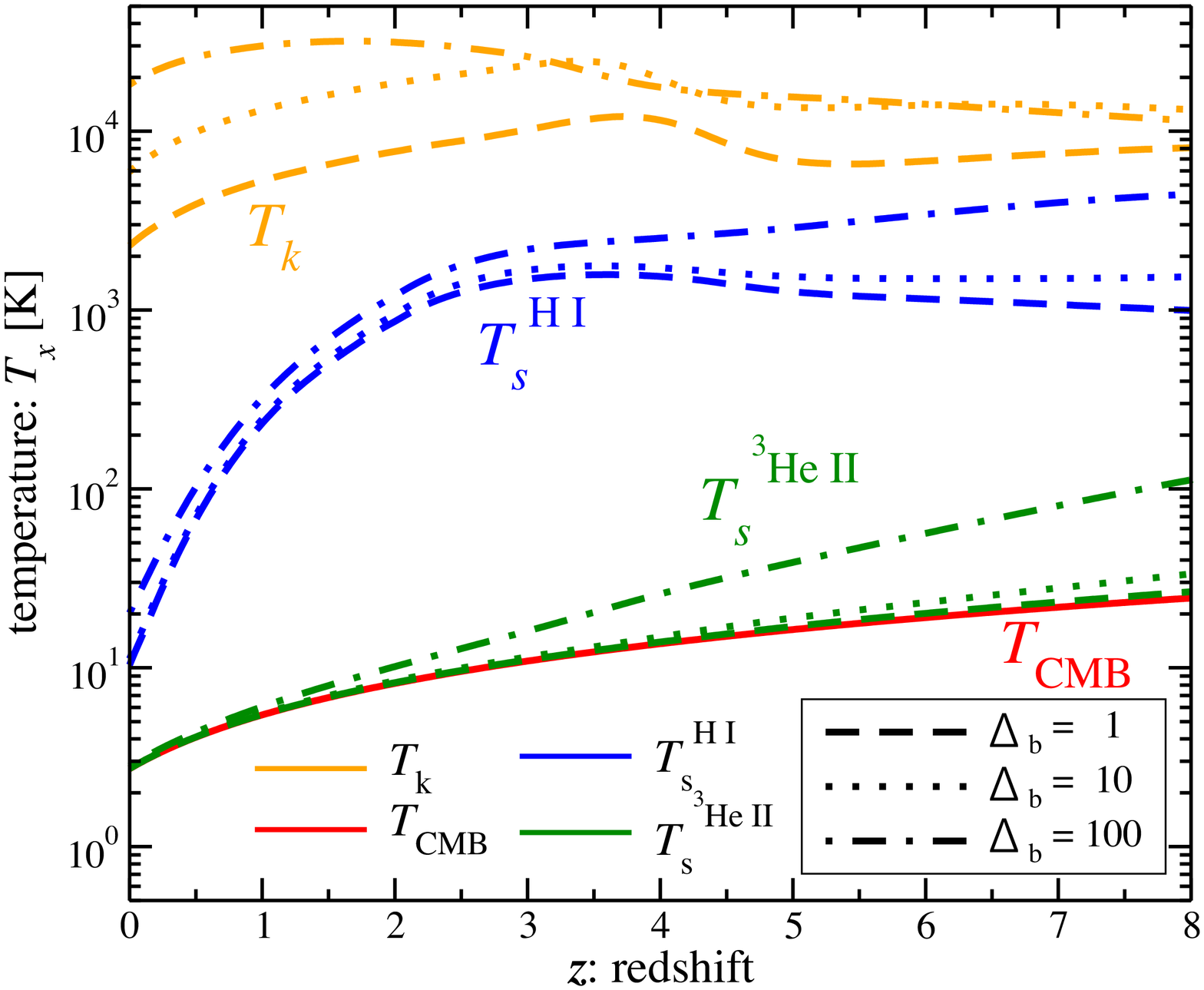}
\hspace{10mm} 
\includegraphics[clip,keepaspectratio=true,width=0.45
  \textwidth]{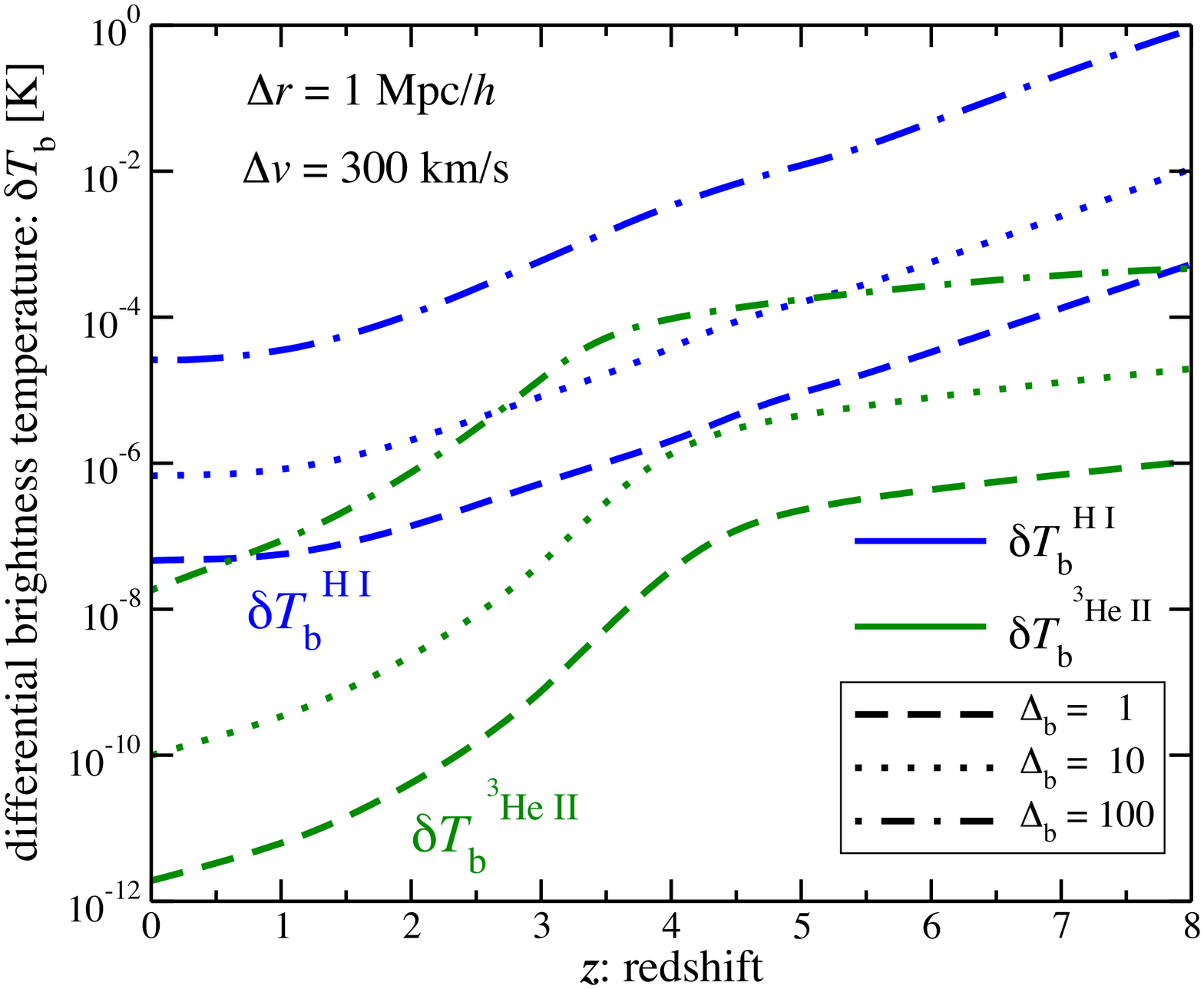}
  \caption[]{(Left) The redshift evolution of the , (from top to
    bottom) kinetic temperature $\Tk$ (orange), the spin temperature
    of \HI\ (blue) and $^3$\HeII\ (green), and the CMB temperature
    $\Tcmb$ (red). (Right) The evolution of the differential
    brightness temperature of \HI\ and $^3$\HeII\ as functions of
    redshift.  In both panels, the different line types represent the
    different values of the baryon density contrast parameter
    $\Delta_{\rm b}$, and the dashed, dotted and dot-dashed lines,
    respectively, correspond to $\Delta_{\rm b} =$ 1, 10 and 100. For
    the estimation of the differential brightness temperature, we
    assume a simple slab model in which the filament is 1 Mpc/$h$ wide
    ($\Delta r=1$ Mpc/$h$) and the proper line-of-sight velocity is
    $\Delta v =300$ km/s.  }
\label{fig:temp}
\end{center}
\end{figure}

The spin temperature is computed with the assumption that the kinetic
temperature $\Tk$ and the color temperature $\Ta$ are coupled to the
gas temperature $\Tg$; i.e. $\Tk \simeq \Ta \simeq \Tg$.  
The evolution of the spin temperature of \HI\ and $^3$\HeII\ are shown
in the left panel of Figure\,\ref{fig:temp}, and the kinetic
temperature ($\Tk$) and the temperature of CMB ($T_{\rm CMB}$) are
also plotted in the same panel.

Although the contribution of the radiative coupling to the spin
temperature is weak at low-redshifts, it becomes important around the
EoR. Moreover some exotic models show that the contribution of X-ray
alters the history of reionization dramatically
\citep{Furlanetto:2006c,Shchekinov:2007}, and uncertainties sill
remain on the estimation of the spin temperature around the EoR and at
even higher-redshift.

\subsection{Differential brightness temperature}
\label{sec:bright}

In radio astronomy, the measured radiation is expressed in terms of
the differential brightness temperature, namely, the difference
between the brightness temperature of the object to that of the CMB;
which is given by
\begin{equation}
  \delta \Tb^{\rm X}(z) \equiv \Tb^{\rm X}(z) - \Tcmb(z) 
  = \frac{[\Ts^{\rm X}(z) - \Tcmb(z)](1 - e^{-\tau_{\rm X}(z)})}{1+z} \, ,
\label{eq:dTb}
\end{equation}
where $\Tb^{X}$ is the brightness temperature and $\tau_{\rm X}(z)$ is
the optical depth, where ${\rm X}$ marks either \HI\ or $^3$\HeII . In
general, the optical depth is \citep{Furlanetto:2006}
\begin{equation}
  \tau_{\rm X}(z) = \frac{g_1}{g_0 + g_1} \frac{c^2 \hp A_{10}}{8\pi \nu_{10}^2 \kB} 
  \frac{n_{\rm X}(z)}{\Ts^{\rm X}(z)} \frac{1}{(dv_{\parallel}/dr_{\parallel})} \, ,
\label{eq:tau}
\end{equation}
where $n_{\rm X}(z)$ is the number density of species ${\rm X}$, and
$dv_{\parallel}/dr_{\parallel}$ is the velocity gradient along the
line-of-sight, including both the Hubble expansion and the peculiar
velocity \citep{Kaiser:1987}.

In the optically thin regime (i.e. $\tau \ll 1$),  
Eq.\,(\ref{eq:dTb}) can be rewritten as 
\begin{eqnarray}
  \delta\Tb^{\rm X}(z) 
  &\simeq& \frac{\left[ \Ts^{\rm X}(z)-\Tcmb(z) \right] \tau_{\rm X}(z)}{1+z} \nonumber \\
  &\simeq& \frac{g_1}{g_0 + g_1} \frac{c^2 \hp A_{10}}{8\pi \nu_{10}^2 \kB} 
  \frac{n_{\rm X}(z)\Delta r}{(1+z)\Delta v}
  \left( 1 - \frac{\Tcmb(z)}{\Ts^{\rm X}(z)} \right) \, ,  
\label{eq:dTb_app}
\end{eqnarray}
where $\Delta r$ denotes the line-of-sight width of the filamentary
structure and $\Delta v$ denotes the proper line-of-sight velocity.
Eq.\,(\ref{eq:dTb_app}) shows that the differential brightness
temperature can be observed in emission(absorption) signal when the
spin temperature is larger(smaller) than the CMB.  In the second line
in Eq.~(\ref{eq:dTb_app}), we used the optical depth given by
Eq.~(\ref{eq:tau}).

\subsection{Signal estimation for a simple slab model}
\label{sec:slab}

In order to calculate the signal we first assume a simple elongated
slab model with constant density.
The filament slab is assumed to have a width $\Delta r = 1$ Mpc/$h$
and the proper line-of-sight velocity $\Delta v=300$ km/s. The
\HI\ column density of this filament corresponds to $N_{\rm HI}=\nHI
\Delta r \simeq 10^{15}-10^{16}$ cm$^{-2}$ with $\Delta_{\rm b}=100$
around the present redshift.
We show the differential brightness temperature of \HI\ and
$^3$\HeII\ in the right panel of Figure \ref{fig:temp} for the
different values of the baryon density contrast parameter,
$\Delta_{\rm b} =$ 1, 10 and 100.

The emission from $^3$\HeII\ shows a different redshift dependence
rerative to that of \HI .
This is because the reionization state of \HI\ is different from that
of He. The drop in $\delta T_{\rm b}^{^3{\rm He II}}$ around $z \sim 4$ 
corresponds to the epoch of \HeII\ reionization 
(\HeII $\rightarrow$\HeIII), where the fraction of \HeII\ 
for the total helium components decreases after that.
Therefore, it is expected that it becomes more difficult to observe
$^3$\HeII\ emission after $z \sim 4$.
However the advantage of the observation with high-frequencies at
low-redshifts also motivates us to attempt to probe the baryons in the
filamentary structures through the emission line of the hyperfine
transition of $^3$\HeII .

\begin{figure}
\begin{center}
\includegraphics[clip,keepaspectratio=true,width=0.7
  \textwidth]{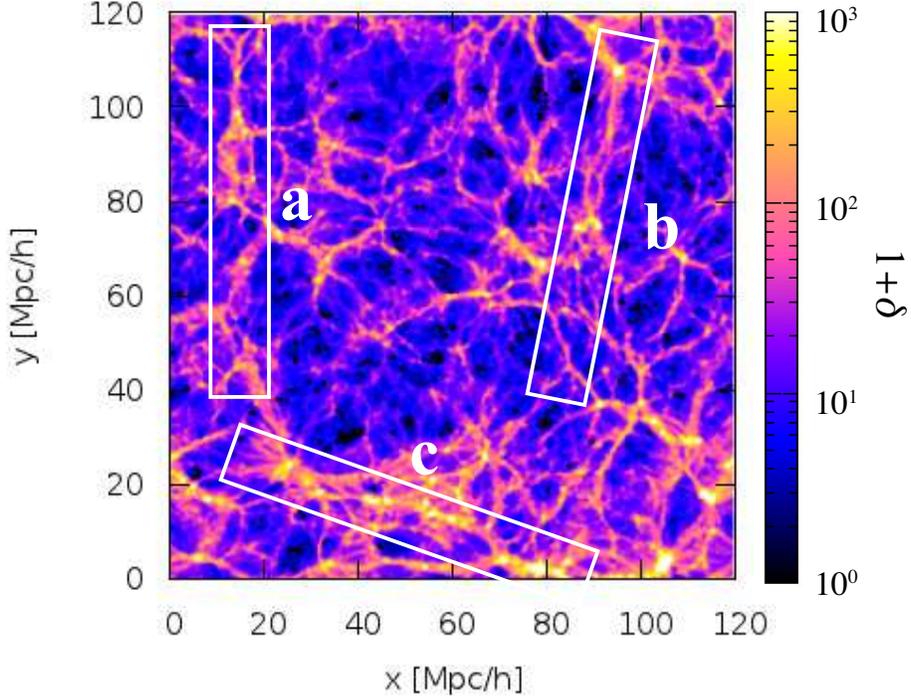}
\vspace{-2mm}
\end{center}
\caption{This figure show a slice of depth 1.4 Mpc/$h$ of the density
  field through the $N$-body simulation at the present redshift. The
  color-bar represents the over density $1+\delta$. The boxes and
  symbols are the appropriately-defined filamentary structures and
  their center positions.
}
\label{fig:delta}
\end{figure}

\section{Signal Estimation from $N$-body simulations}
\label{sec:signal}

In the previous section, we calculated the brightness temperature of
the hyperfine emission assuming the overdensity $\delta$, the proper
line-of-sight velocity $\Delta v$ and the width $\Delta r$ of the
filament.  As a next step, we carry out the same calculation but with
filaments taken from a cosmological $N$-body for the large-scale
structure.  In this section the $N$-body simulations used to make the
large-scale structures and the procedure to estimate the signal of
hyperfine transition from them are described.  Furthermore, a number
of elongated filamentary structure and their observed signal as seen
from various viewing angles are considered.

\subsection{$N$-body Simulation}
\label{sec:sim}

To obtain the $N$-body $512^3$ particles simulation a parallel
Tree-Particle Mesh code $Gadget$-2 \citep{Springel:2005} in its full
Tree-PM mode was run. The simulation size is $120$ Mpc/$h$ on a side,
the minimum mass resolution corresponds to $9.68 \times 10^8
M_{\odot}/h$ and, as mentioned before, adopts the WMAP-7yrs
cosmological model.  The matter density field estimated through the
$N$-body simulation at the present time is shown in
Figure~\ref{fig:delta}. The frames and the symbols represent the
appropriately-determined filamentary structures and their center
positions.

We here estimate the signal of the hyperfine transition for \HI\ and
$^3$\HeII\ based on the $N$-body simulation of the large-scale
structures.
To calculate the brightness temperature, we first divide the
simulation box into $256^3$ grids and estimate the density contrast
$\delta \equiv (\rho-\bar{\rho})/\bar{\rho}$ and the proper
line-of-sight velocity $\Delta v$ with Cloud-in-Cell (CIC)
interpolation on each grid.  Then the differential brightness
temperature is calculated through Eq.\,(\ref{eq:dTb}) adopting the
values of the density contrast and the proper line-of-sight velocity
on each grid. Both signals from \HI\ and $^3$\HeII\ are evaluated
according to the description in Sec.~\ref{sec:model}.

\begin{figure}
\begin{center}
\includegraphics[keepaspectratio=true,width=0.75
  \textwidth]{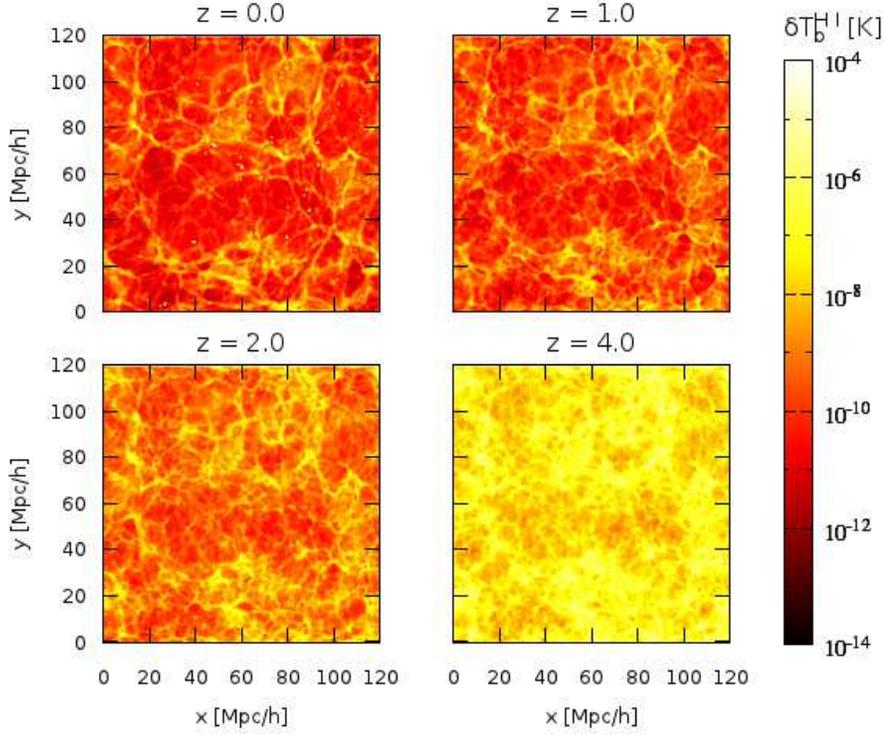}
\end{center}
\caption{Maps of the differential brightness temperature of \HI\ at at
  $z=0$ (Top-Left), $z=1$ (Top-Right), $z=2$ (Bottom-Left) and $z=4$
  (Bottom-Right).  We take the density contrast and the proper
  line-of-sight velocity from the snapshots of the $N$-body simulation
  and then calculate the differential brightness temperature through
  Eq.\,(\ref{eq:dTb}).  All slices are 120 Mpc/$h$ on the side and the
  depth of 1.4 Mpc/$h$.  }
\label{fig:maps}
\end{figure}

\begin{figure}
\begin{center}
\includegraphics[keepaspectratio=true,width=0.75
  \textwidth]{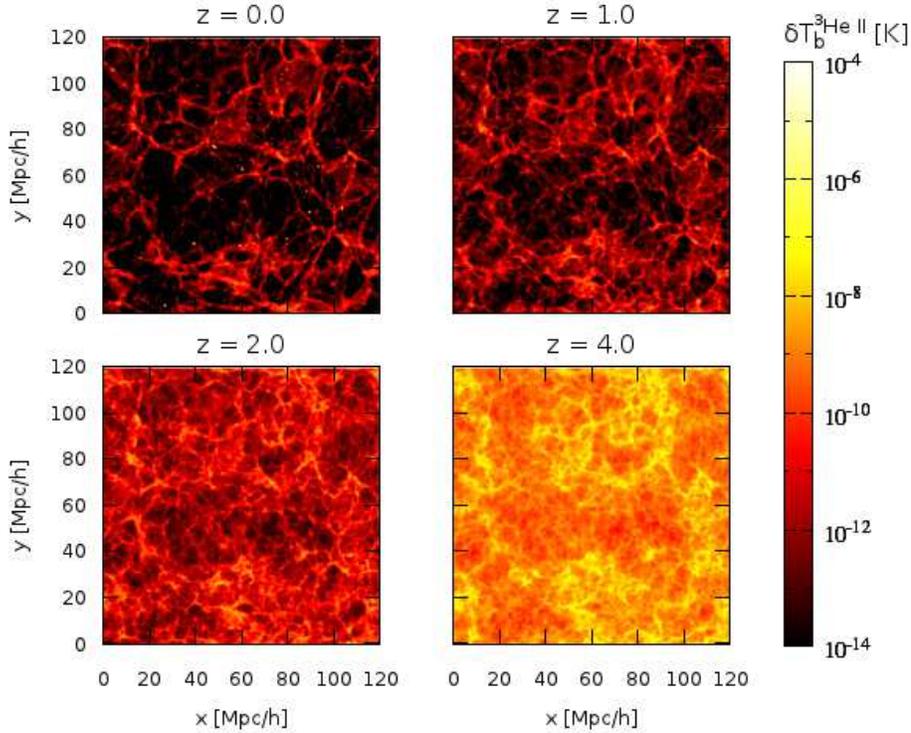}
\end{center}
\caption{Same as Figure~\ref{fig:maps}, except the maps are for the
  differential brightness temperature of $^3$\HeII. }
\label{fig:maps_3he}
\end{figure}

Figures~\ref{fig:maps} and~\ref{fig:maps_3he} show the differential
brightness temperature of thin slices across the simulation box at
different redshifts for \HI\ and $^3$\HeII\, respectively. 
The simulation outputs are shown at $z=0$, 1, 2 and 4. The depth of
each slice is 1.4 Mpc/$h$ in all cases, where the two-dimensional maps
shown in the figures are integrated over the depth.  The depth of 1.4
Mpc/$h$ corresponds to the frequency bandwidth of $\sim$0.7 MHz for
the \HI\ survey and $\sim 4$ MHz for the $^3$\HeII\ survey at $z=0$.

The filamentary structures in both cases is seen through the
brightness temperature, where the amplitudes are $\dTb~\sim~10^{-6}$ K
for \HI\ and $\sim~10^{-9}$ K for $^3$\HeII\ in the maps at $z=0$. The
signals are lower than the case we calculated earlier, assuming a slab
filamentary structure with constant baryon overdensity $\Delta_{\rm b}
\simeq 100$ (see the right panel of Figure~\ref{fig:temp}) -- this
also applies for higher redshifts.
One can clearly see from Figure~\ref{fig:delta} that at present time
the density contrast of the filaments is $\Delta_{\rm b} \la 100$,
which explains the lower values we obtain.  We can also clearly see
the drastic change in the intensity of $^3$\HeII\ from $z=4$ to $z=2$,
which reflects the \HeII\ reionization.

\subsection{Signal from a filamentary structure}

Filaments can extend to very large distances reaching many tens of
mega-parsecs, as seen in simulations and in real observations
\citep{Erdogdu:2004,Choi:2010,Pandey:2011}.  
When a filament is elongated along the line-of-sight direction one
expects to detect the highest signal from the filamentary
structure. Conversely, when the filament and the line-of-sight
direction are perpendicular to each other, the signal should be lower
than the former case. Therefore, the signal from a filamentary
structure depends on the viewing angle relative to the line-of-sight
direction.

\begin{figure}
\begin{center}
\includegraphics[clip,keepaspectratio=true,width=0.40
  \textwidth]{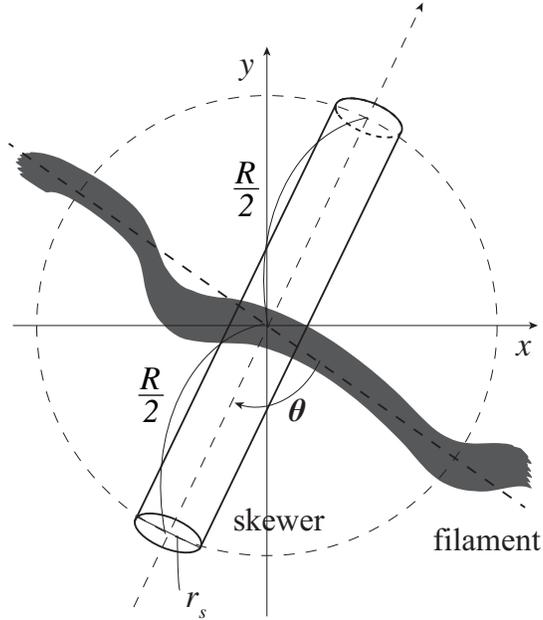}
\end{center}
\caption{The positional relation between the filamentary structures
  and the observational skewer. Here, $\theta$ represents the angle
  between the skewer and long axes of the filament's. $R$ and $r_s$
  represent the length along the line-of-sight and the spacial
  resolution of the skewer. The line-of-sight and of the filament's
  long axes direction define the x-y plane. }
\label{fig:skewer}
\end{figure}

The brightness temperature, $\dTb$ of the three filaments is shown in
Figure~\ref{fig:delta} and are labeled with ``a", ``b" and ``c".
We take the signal within a cylindrical region that passes along the
line of sight, which we refer to as the observational skewer.  The
cross section of the skewer represents the telescopes beam, i.e.,
resolution. The skewer is assumed to be on the $x$-$y$ plane for each
filament shown in Figure~\ref{fig:delta} and rotate a skewer by an
angle $\theta$ around the center of the filament. The thickness of the
skewer corresponds to the spatial resolution with which one observes.
The positional relation between the filamentary structures and the
observed skewer is shown in Figure~\ref{fig:skewer}.  The cartoon is
shown on the plane that is define by the filament and the
line-of-sight directions.
Here $\theta$ represents the angle between the filament and the
obvervational skewer. The $\theta=0^\circ$ is defined such that the
line-of-sight direction is parallel to the filament long axis.
The signal at zero redshift as a function of the viewing angle for
\HI\ is shown in Figure~\ref{fig:filaments_HI} and for $^3$\HeII\ in
Figure~\ref{fig:filaments_3HeII}.  In these figures, the results for
the different values of the spacial resolution, $r_s = 1$, $3$, $5$
and $10$ Mpc/$h$ is shown, and the length of the skewer is $R=100$
Mpc/$h$ for all cases.

We should note that these results have no contribution from the
high-density region whose density contrast is $\delta \ge 200$ because
we masked the signal from such regions to eliminate the contributions
from galaxies.
On the other hand, the haloes with small mass which is often called as
minihalo may also contributes to signal along the line of sight, but
our simulation does not have enough resolution to resolve such haloes.

The small spikes seen in Figures~\ref{fig:filaments_HI} and
\ref{fig:filaments_3HeII} reflect the contributions from the other
structures around the filament. Compared to filaments a and b,
filament-c has more widespread structure which explains the weak
angular dependence of filament-c.
It is clear that if one observes a filament, which extends along the
line-of-sight, the amplitude of the expected emission signal is
stronger. For example, in the case of filament-a, $\dTb \sim 10^{-5}$
K for \HI\ and $\sim 10^{-8}$ K for $^3$\HeII\ assuming spatial
resolution of $r_s=1$ Mpc/$h$, which are more than 10 times larger
than the case in which the two directions are perpendicular.
Furthermore, for coarser resolution (thicker skewer), the signal is
also expected to be higher.  In the following sections, the prospects
for observing the signals of the hyperfine transition are discussed.

\section{Prospect of observation}
\label{sec:obs}

\subsection{Observational sensitivity}
\label{sec:sensitivity}

In this subsection, we summarize the sensitivity of relevant
radio-telescope to estimate the prospects for observing the signals of
\HI\ and $^3$\HeII . Here we consider single dish and interferometric
radio telescopes, and for each case we estimate the noise level
expected, using the expressions of \citet{Furlanetto:2006}.

The resolution of the radio-telescope can not exceed the diffraction
limit, which depends on the largest dimension of the telescope 
$D_{\rm max}$ and the observing frequency $\lambda$ as 
$\theta_{\rm D}=\lambda/D_{\rm max}$.

The sensitivity or the brightness temperature uncertainty of radio
telescope is
\begin{equation}
  \delta T_{\rm N}(\lambda) 
  \simeq \frac{\lambda^2/\theta_{\rm s}^2}{A_{\rm tot}} 
  \frac{T_{\rm sys}}{\sqrt{\Delta \nu t_{\rm obs}}} \, ,
\label{eq:dtn}
\end{equation}
where 
$\lambda$ is the observing frequency, 
$\theta_{\rm s}$ is the angular resolution, 
$T_{\rm sys}$ is the system temperature, 
$A_{\rm tot}$ is the total collecting area,
$\Delta \nu$ is the band width, 
and
$t_{\rm obs}$ is the integration time of the observation.

The system temperature is given by 
\begin{equation}
T_{\rm sys} \equiv T_{\rm sky} + T_{\rm rec},
\label{eq:tsys}
\end{equation}
where the first term is due to the sky itself and is dominant at low
frequencies ($\nu \lsim 150$MHz), whereas the second is due to the
receiver and dominates at high frequencies ($\nu \gsim 150$MHz).

The sky temperature at high-latitude is roughly given as
\begin{equation}
  T_{\rm sky} \sim 180 \left( {\nu}/{180 {\rm MHz}} \right)^{-2.6} {\rm K} \, \,.
\label{eq:tsky}
\end{equation}
Therefore,  the noise of the interferometer due to the sky can be
written using Eq.\,(\ref{eq:dtn}) as,
\begin{equation}
  \delta T_{\rm N}^{\rm sky}(\nu) \simeq 5.0 \, \epsilon_{\rm ap}^{-1}
  \left( \frac{10^5\;{\rm m}^2}{A_{\rm tot}} \right)
  \left( \frac{1'}{\Delta \theta} \right)^2
  \left( \frac{1+z}{1} \right)^{4.6}
  \left( \frac{1420\,{\rm MHz}}{\nu_0} \right)^{4.6}
  \left( \frac{\rm MHz}{\Delta \nu} \frac{100\;{\rm hr}}{t_{\rm obs}} \right)^{1/2} 
       \ \mu{\rm K} \, , 
\label{eq:noise_sky}
\end{equation}
where 
$\epsilon_{\rm ap}$ is the aperture efficiency of a dish, 
$\nu$ is the observational frequency, 
$\Delta \theta$ is the beam angular resolution,
and 
$\nu_0$ represents the rest frame frequency of the observed hyperfine
transition; i.e., $\nu_0$=1420 MHz for \HI\ or $\nu_0$=8667 MHz for
$^3$\HeII . 

Similarly, the noise due to the receiver, which is roughly frequency
independent, can be written from Eq.\,(\ref{eq:dtn});
\begin{equation}
  \delta T_{\rm N}^{\rm rec} \simeq 23.9 \, \epsilon_{\rm ap}^{-1}
  \left( \frac{T_{\rm rec}}{30\;{\rm K}} \right)
  \left( \frac{10^5\;{\rm m}^2}{A_{\rm tot}} \right)
  \left( \frac{1'}{\Delta \theta} \right)^2
  \left( \frac{1+z}{1} \right)^{2.0}
  \left( \frac{1420\,{\rm MHz}}{\nu_0} \right)^{2.0}
  \left( \frac{\rm MHz}{\Delta \nu} \frac{100\;{\rm hr}}{t_{\rm obs}} \right)^{1/2} 
       \ \mu{\rm K} \, , 
\label{eq:noise_rec}
\end{equation}
where $T_{\rm rec}$ is the receiver noise.  

The total noise is given by combining Eqs.~(\ref{eq:noise_sky}) and
(\ref{eq:noise_rec}) as
\begin{equation}
  \delta T_{\rm N} = 
  \left(
  \delta T_{\rm N}^{\rm sky} + \delta T_{\rm N}^{\rm rec}
  \right) \times  \left\{
 \begin{array}{lcl}
   1  & & ({\rm single ~ dish}) \\
   1/\sqrt{N_{\rm B}} & & ({\rm interferometer})   ,
 \end{array}
 \right.
\label{eq:noise_inter}
\end{equation}
where $N_{\rm B}=N_{\rm dish}(N_{\rm dish}-1)/2$ is the number of
pair-wise correlations or base lines, and $N_{\rm dish}$ is the number
of interferometeric elements in the telescope.  The noise for the
interferometer can be reduced by a factor of $1/\sqrt{N_{\rm B}}$.

We consider the sensitivity of relevant current and future radio-telescopes:
the Green Bank Telescope (GBT), 
the Arecibo Radio Telescope (Arecibo), 
the Five-hundred-meter Aperture Spherical Telescope (FAST), 
the Giant Metrewave Radio Telescope (GMRT), 
the Expanded VLA (EVLA), 
MeerKAT 
and 
the Square-Kilometer Array (SKA).
The characteristic parameters for these radio telescopes are shown in
Table~\ref{tb:survey}. We should note that the operating range of GMRT
does not have enough frequency range to observe the signal of
$^3$\HeII\ at low-redshifts and its achievable redshift range is $z >
4.7$.  The operating range of FAST also does not reach the restframe
frequency of $^3$\HeII. However,  future upgrade may allow to
observe up to 8 GHz \citep{Nan:2011}.

\begin{table}
\begin{center}
\begin{tabular}{lrrrcrrrrr}
\hline \hline
Telescope 
& \makebox[15mm][r]{$R_{\rm dish}$ [{\rm m}]}
& \makebox[15mm][r]{$N_{\rm dish}$}   
& \makebox[20mm][r]{$A_{\rm tot}$ [${\rm m}^2$]}
& \makebox[15mm][c]{$\epsilon_{\rm ap}$} 
& \multicolumn{3}{c}{Operating range } 
& \makebox[20mm][r]{$D_{\rm max}$ [km]} 
\\
\hline 
{\bf Single dish}\\
\hspace{3mm} GBT 
  & 100 & --- &  7,850 & 0.7  &    100 MHz & - & 116 GHz & 0.1 \\
\hspace{3mm} Arecibo 
  & 305 & --- & 73,000 & 0.8  &     47 MHz & - &  10 GHz & 0.3 
\\
\hspace{3mm} FAST 
  & 500 & --- & 200,000   & 0.8  &  70 MHz & - &  3 GHz  & 0.5 \\
\hdashline
{\bf Interferometers}\\
\hspace{3mm} EVLA 
  & 25  & 27 &  13,300   & 0.8  &  1,000 MHz & - &  50 GHz  & 1\ -\ 36\\
\hspace{3mm} GMRT
  & 45  & 30 &  60,750   & 0.8  &     50 MHz & - & 1.5 GHz  & 25 \\
\hspace{3mm} MeerKAT (phase 2) 
  & 13.5  & 64 & 9,160   & 0.8  &     580 MHz & - &  14.5 GHz  & 20 \\
\hspace{3mm} SKA (phase 2) 
  & 15  & 1,500 & 300,000   & 0.8  &     70 MHz & - &  10 GHz  & 5c \\
\hline \hline
\end{tabular}
\caption{The characteristic parameters for the specification of the
  radio-telescopes. GBT and Arecibo are the single dish telescopes and
  the others are the radio-interferometers. $R_{\rm dish}$ is the
  diameter of a dish for one telescope, $N_{\rm dish}$ is the number
  of telescopes, and $A_{\rm tot}$ is the effective total collecting
  area of the telescope. The parameter $\epsilon_{\rm ap}$ represents
  the aperture efficiency of a dish. $D_{\rm max}$ is the maximum base
  line length of the telescope; for EVLA the value depends on the
  configuration of the telescopes, and for SKA we take the extent of
  compact core.
}
\label{tb:survey}
\end{center}
\end{table}

The \HI\ and $^3$\HeII\ differential brightness temperature and the
sensitivities as a function of redshift for the slab model case are
shown in Figures~\ref{fig:dtb} and \ref{fig:dtb_3He}, respectively.
The signal curves are the same as the ones shown in the right panel of
Figure~\ref{fig:temp}, whereas the solid curves in this figure show
the sensitivity of the various telescopes.
Figures~\ref{fig:dtb} and \ref{fig:dtb_3He} show three types of
observation with different angular resolutions: $\Delta \theta = $30,
10 and 1 arcmin.  The system temperature includes the sky and receiver
temperatures, where we assume that $T_{\rm rec}=30$~K and 
$T_{\rm sky}$ is taken from Eq.\,(\ref{eq:tsky}).
The sensitivity is given assuming 100 hours of integration time and a
bandwidth $\Delta \nu=30$ MHz for \HI\ and $\Delta \nu=$200 MHz for
$^3$\HeII. These values for the bandwidth correspond to $\sim 100$
Mpc/$h$ depth along the line-of-sight around $z=0.5$.  We should note
that, for our case, the sensitivity of EVLA and FAST are almost the
same, hence, we only plot the sensitivity curve for FAST.
The relationship between the comoving distance and the bandwidth is
given in Appendix~\ref{sec:size}.

\begin{figure}
\begin{center}
\includegraphics[clip,keepaspectratio=true,width=0.7
  \textwidth]{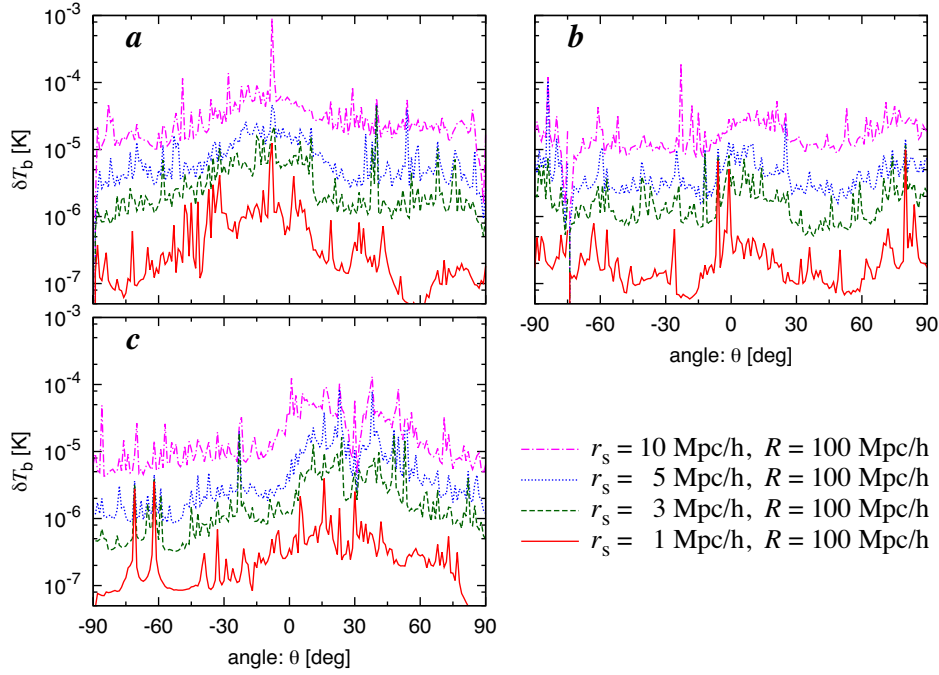}
\end{center}
\caption{The signal of \HI\ from the filamentary structures at
  $z=0$. The angle $\theta$ represents the angular size between a
  filament and the line-of-sight, $r_s$ and $R$ represent the spacial
  size and the line-of-sight width for a skewer. The different color
  lines show the different values of $r_s$; which are 10 Mpc/$h$
  (magenta), 5 Mpc/$h$ (blue), 3 Mpc/$h$ (green) and 1 Mpc/$h$ (red),
  and the width is fixed to be $R=100$ Mpc/$h$ for all cases. The
  positional relation between a filament and a skewer is pictured in
  Figure~\ref{fig:skewer}. The label ``a", ``b" and ``c" correspond to
  the filaments shown in Figure~\ref{fig:delta}.}
\label{fig:filaments_HI}
\end{figure}

\begin{figure}
\begin{center}
\includegraphics[clip,keepaspectratio=true,width=0.7
  \textwidth]{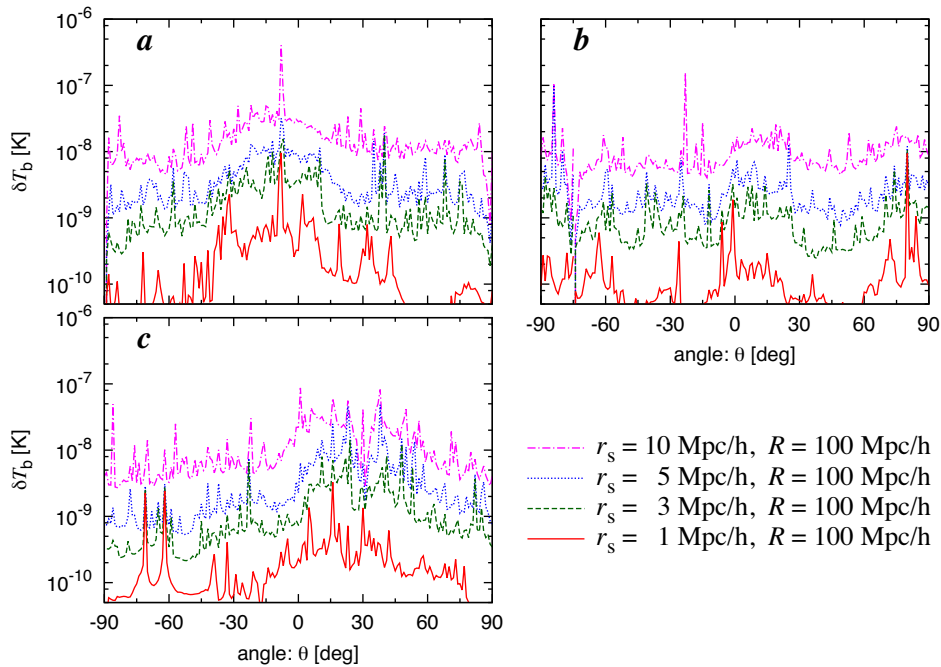}
\end{center}
\caption{Same as Figure~\ref{fig:filaments_HI}, but for the signals of $^3$\HeII .}
\label{fig:filaments_3HeII}
\end{figure}

\begin{figure}
\begin{center}
\includegraphics[clip,keepaspectratio=true,width=0.8
  \textwidth]{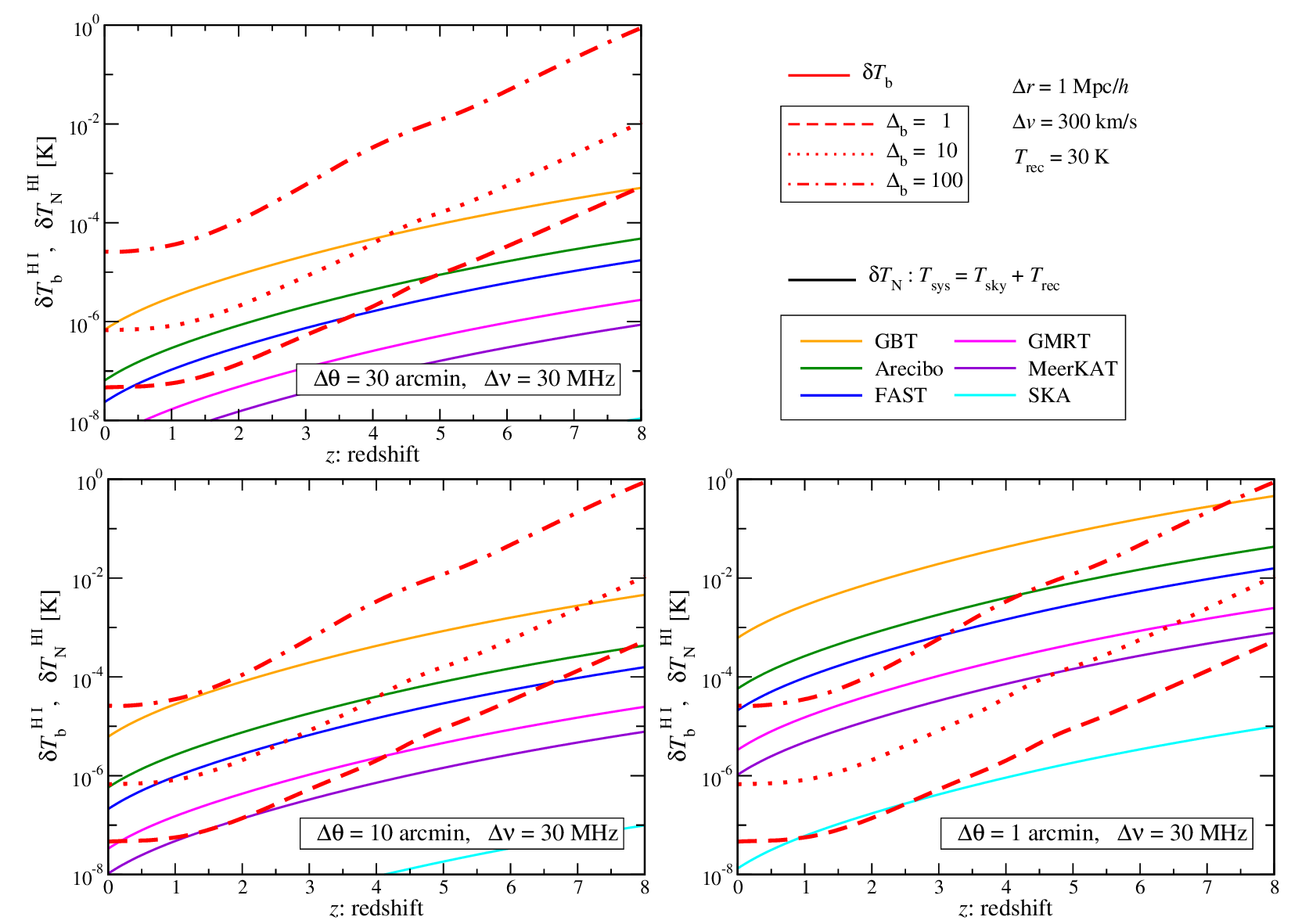}
\caption{The signal and noise spectrum for the hyperfine transition of
  \HI\ as a function of redshift. The signal represents the
  differential brightness temperature of \HI\ (red-lines), which is
  same as shown in the right panel of Figure~\ref{fig:temp}. The noise
  spectrum is adopted for the observation at the rest frame frequency
  of 1.4 GHz with GBT (orange), Arecibo (green), EVLA (violet), FAST
  (blue), MeerKAT(violet) and SKA (cyan), respectively.  
  The solid curves represent the noise taking into account only the
  sky temperature, and the dashed curves with same color represent the
  noise taking into account both the sky temperature and the receiver
  noise of $T_{\rm rec} = 30$ K.  The three panels show the difference
  of the noise spectra with the different values of the spacial
  resolution $\Delta \theta$ for the observation; we adopt the values
  to be
  $\Delta \theta = 30$ arcmin (Top-Left), 
  $\Delta \theta = 10$ arcmin (Bottom-Left), and 
  $\Delta \theta = 1$ arcmin (Bottom-Right), respectively. 
  The frequency band width of $\Delta \nu = 30$ MHz, 
  which corresponds to $\sim$100 Mpc/$h$ width along the line-of-sight 
  at $z=0.5$, is adopted for all cases. 
  We assume the 100 hours of integration time for all telescopes. 
 }
\label{fig:dtb}
\label{fig:dtb_HI}
\end{center}
\end{figure}

\begin{figure}
\begin{center}
\includegraphics[clip,keepaspectratio=true,width=0.8
  \textwidth]{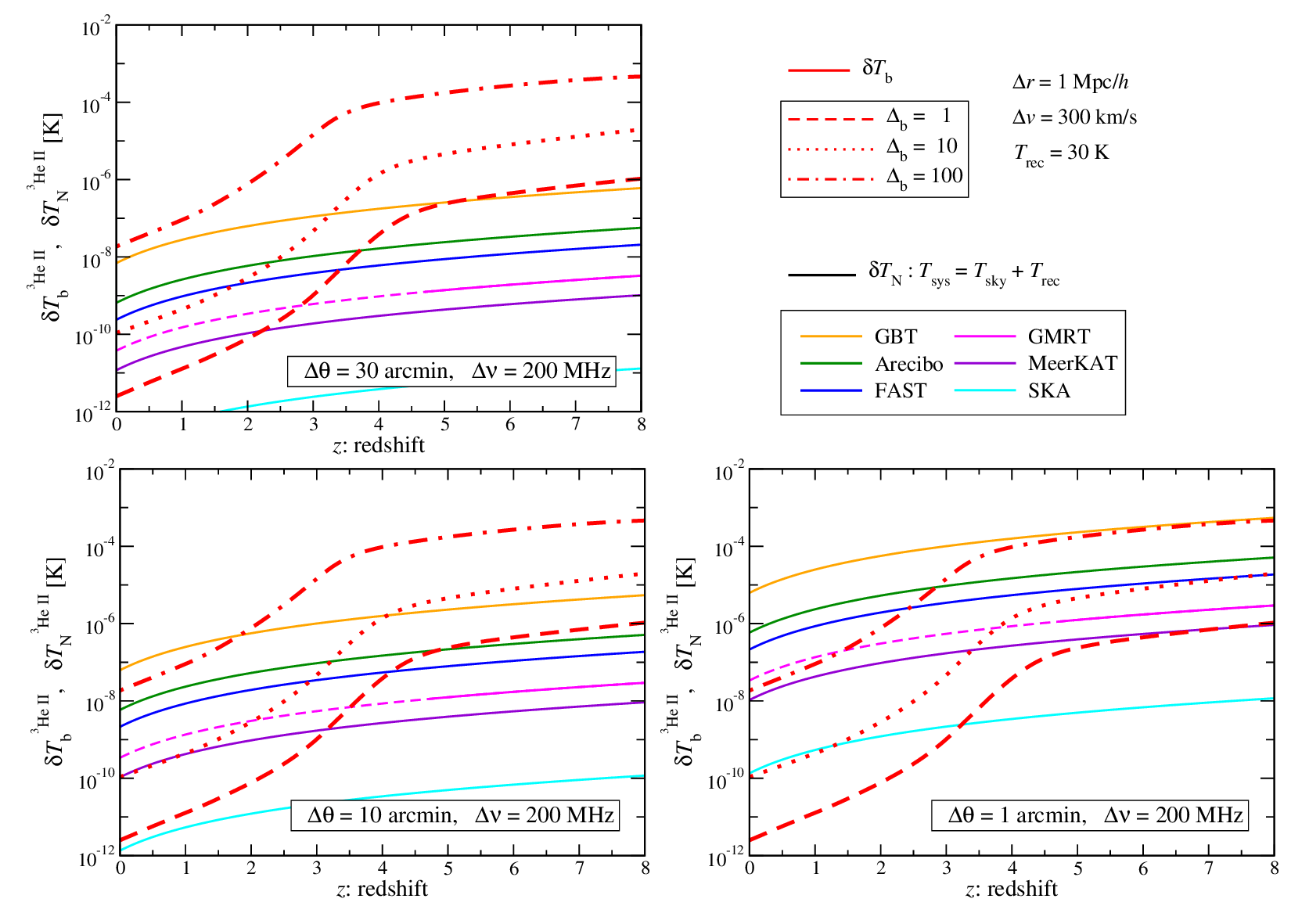}
  \caption{Same as the right panel in Figure~\ref{fig:dtb_HI}, except
    the signals are the differential brightness temperature of
    $^3$\HeII\ and the noise spectra are adopted for the observation at
    the rest frame frequency of 8.7 GHz. The frequency band width of 
    $\Delta \nu = 200$ MHz, which corresponds to $\sim$100 Mpc/$h$ width 
    along the line-of-sight at $z=0.5$, is adopted for all cases.
}
\label{fig:dtb_3He}
\end{center}
\end{figure}

\begin{figure}
\begin{center}
\includegraphics[clip,keepaspectratio=true,width=0.9
  \textwidth]{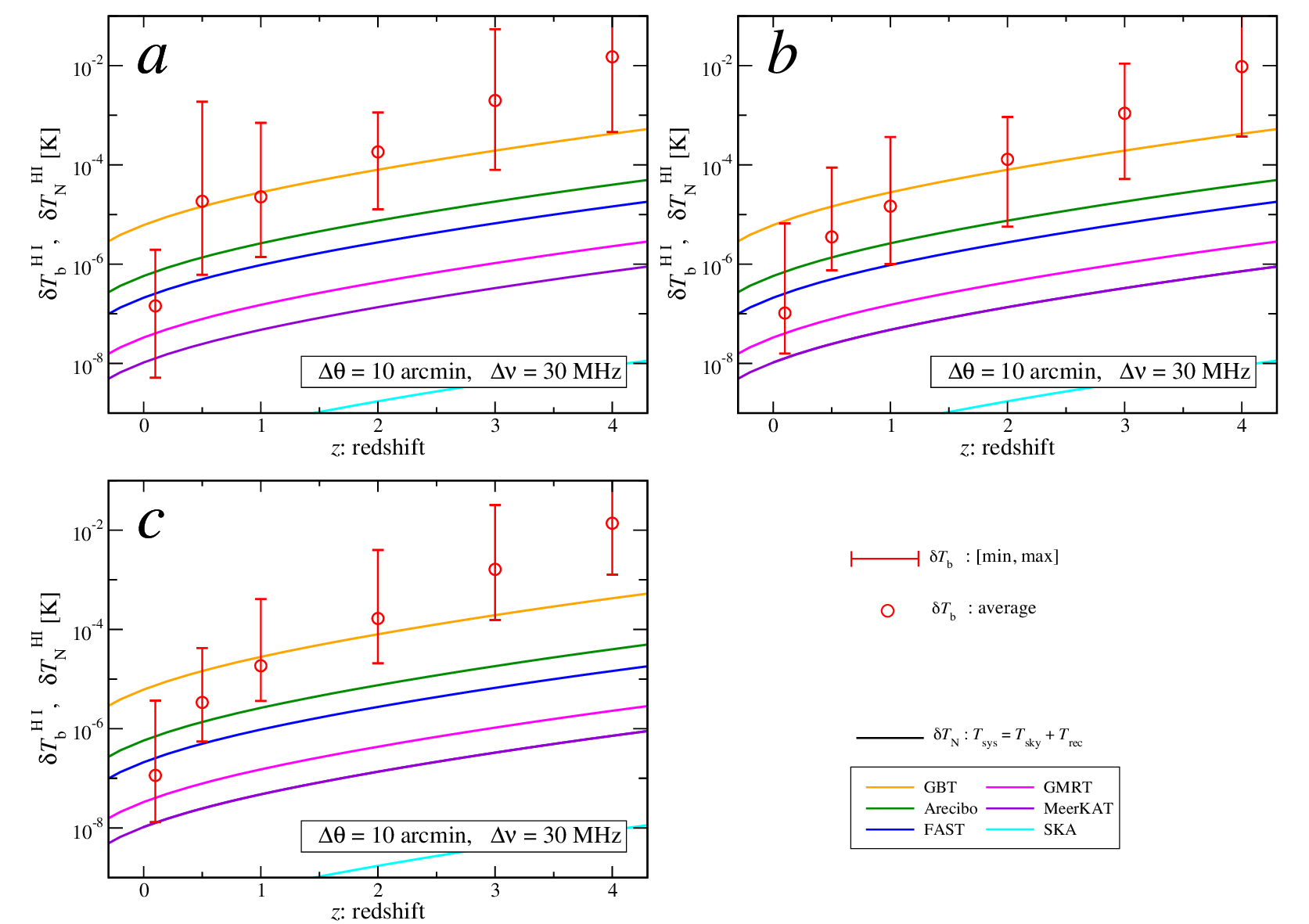}
\end{center}
\caption{The redshift evolution of the \HI\ signals from the
  filamentary structures labeled as a, b and c, and we show the
  results at $z=0.1$, $0.5$, $1$, $2$ and $4$. Each data point represents the
  range of signal between maximum and minimum amplitudes, and the
  symbol represents the average value. 
}
\label{fig:filaments_HI_z}
\end{figure}

\begin{figure}
\begin{center}
\includegraphics[clip,keepaspectratio=true,width=0.9
  \textwidth]{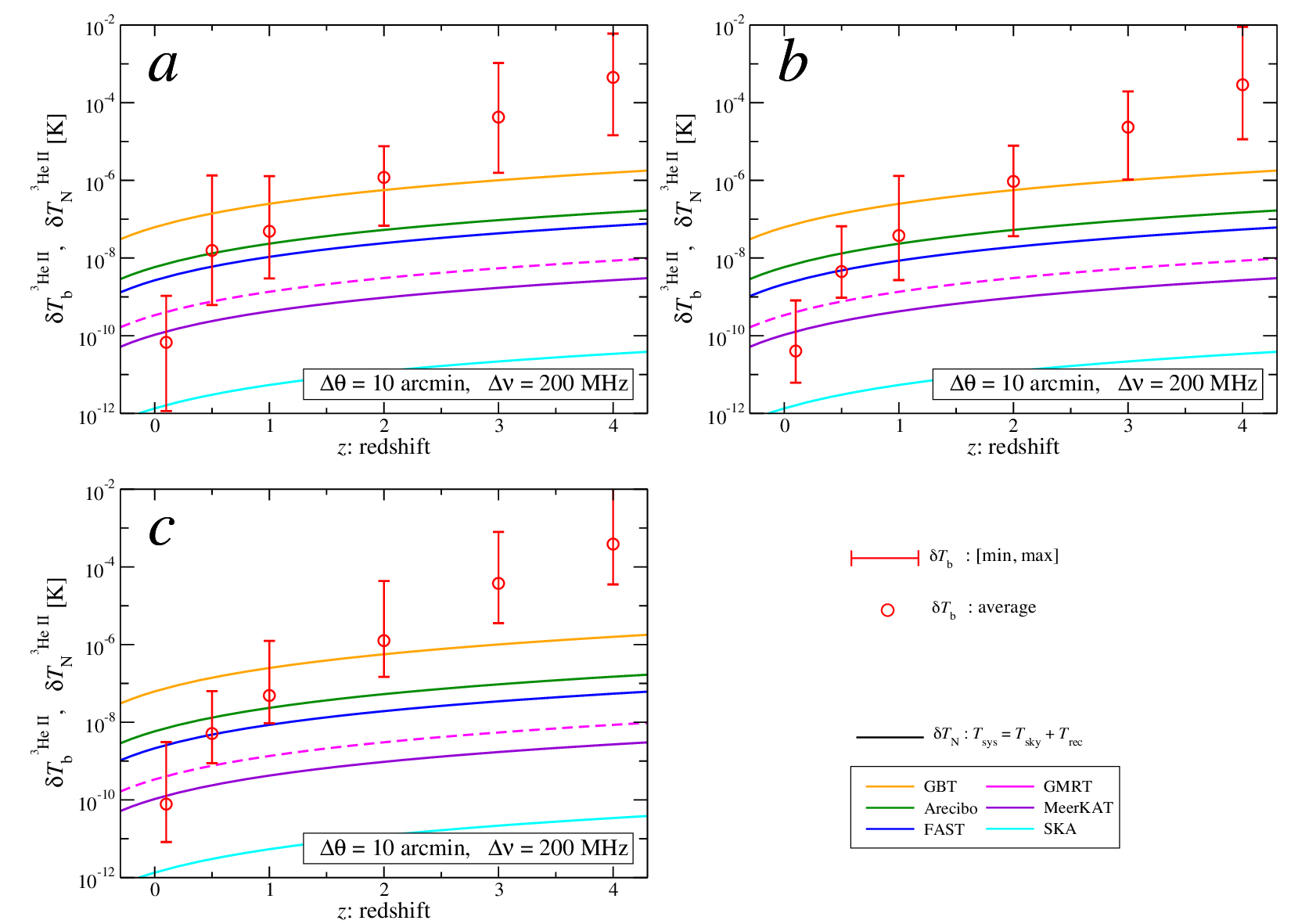}
\end{center}
\caption{Same as Figure~\ref{fig:filaments_HI_z}, but for the signal
  and sensitivity curves of $^3$\HeII .}
\label{fig:filaments_He_z}
\end{figure}

As shown in Figures~\ref{fig:maps} and \ref{fig:maps_3he} for the case
of the $N$-body simulations, the emission signal is expected to
increase at higher redshifts for both \HI\ and $^3$\HeII .  
In Figures~\ref{fig:filaments_HI_z} and \ref{fig:filaments_He_z}, we
show the signal for the $N$-body simulations case again.  The
brightness temperature is shown for the three filamentary structures
(a, b and c) at redshifts, $z=(0.1,0.5,1,2,4)$, for \HI\ and
$^3$\HeII, respectively, assuming an angular resolution of
$10$~arcmin. We also plot the sensitivity curves for each survey in
the same panels. The sensitivity is calculated in the same way as in
Figures~\ref{fig:filaments_HI} and \ref{fig:filaments_3HeII}.  
The points show the brightness temperature for each filament at a
given redshift averaged over the viewing angles. The bar represent and
range of values of the observed temperature for the range of viewing
angles, where the maximum is obtained at viewing angle of $0^\circ$
and the minimum at viewing angle of $90^\circ$.

Figure~\ref{fig:filaments_HI_z} shows that for \HI, the amplitude of
signal is significantly larger than that of \iHeII.  If we focus on
the redshift range of $z\sim 0- 1$, the brightness temperature is
$\dTb \sim 10^{-6}$ to $10^{-3}$ K with angular resolution of
10~arcmin, which corresponds to $\sim$ 5.5 comoving Mpc/$h$ at
$z=0.5$.  The figure clearly shows that such filaments are in
principle detectable with current radio telescopes up to high
redshifts. The beam size of the present receiver for this frequency
range is adequate; e.g., $9$ arcmin for GBT and $\sim 3.5$ arcmin for
Arecibo.  We assume here 100 hours integration time. Note however that
GMRT, EVLA and future telescopes may have enough sensitivity to detect
the signal even with a shorter integration time.

On the other hand, compared to the signal of \HI , the signal of
$^3$\HeII\ is much weaker and the amplitude is on the order of
micro-kelvin at $z\sim 0-1$ as shown in Figure~\ref{fig:filaments_He_z}. 
The sensitivity is calculated assuming $10$~arcmin resolution and 100
hours of observation time. Such a signal is very hard to detect by the
currently available telescopes, maybe with the exception of GMRT.
GMRT is the most suitable of the currently operational telescopes for
the observation of $^3$\HeII.
Unfortunately, its operating frequency range is less than 1.5 GHz
which is suitable for detecting $^3$\HeII\ only at $z > 4.7$. Future
telescopes however, will be able to detect such feeble emission after
100 hours of observation.

Obviously, the beam size at the frequencies relevant to \iHeII\ is
smaller than that of the beam for \HI\, due to the higher frequency of
the transition.  Therefore, the field of view for the observation is
much smaller.  In general, the field-of-view for the single dish
telescope corresponds to the beam size, and the observing area are
typically small.
Due to the limited size of the beam relative to that of the objects of
interest (filaments), the integrated sensitivity at high frequency is
small, hence, a single dish telescope seems unsuitable for the
detection of the $^3$\HeII\ signal at low-redshifts.
In comparison, the field-of-view of an interferometer is
given by the dish size of a small telescope; $\theta_{\rm FoV} \simeq
\lambda/R_{\rm dish}$. 
Therefore, interferometers typically have a larger field-of-view than
the single dish telescopes, and one can obtain a larger skewer cross
section, e.g. $\theta_{\rm FoV} \sim 4$ arcmin for EVLA and
$\theta_{\rm FoV} \sim 8$ arcmin for MeerKAT.  
Future telescopes such as MeerKAT and SKA have enough sensitivity for
detecting the $^3$\HeII\ signal around these redshifts.

Furthermore, at higher redshift, $z \ge 2$, we still have possibility
for detecting the $^3$\HeII\ signal with present instruments such as
Arecibo and EVLA.
The angular size increases at higher redshifts, and the observing
frequency decreases. The beam size of Arecibo reaches $3$ arcmin at 
$z = 4$ and it corresponds to the spacial size of 
$r_s \sim 5~\mathrm{Mpc/h}$.
Around these redshifts, we can not use the galaxy catalogs of the
optical galaxy redshift survey, but the results from the observation
of Ly-$\alpha$ can be adopted as tracers to find the high-density
regions.

\subsection{Observational Issues}
\label{sec:obs}

\subsubsection{Observational Strategy:}  
The strategy for choosing the observing area and frequency is a key
aspect for the detection of filaments with \HI\ and \HeII.  Obviously,
for a highly sensitive instrument, like SKA, one can choose to survey
large area and see whether there is any localized excess in the
intensity of \HI\ and \HeII\ which will imply large scale structure
with excess neutral fraction of these specs.
  
The sensitivity of other instruments is not as good as SKA, therefore
one has to adopt a different strategy for such a detection (this is
also true for the initial and even intermediate stages of SKA.
Therefore, for such instruments we assume some prior knowledge on the
location of filaments. Namely, we use information from other
observations would provide the necessary information on the location
and orientation of filaments.  Recent galaxy redshift surveys such as
2dF and SDSS have revealed the vast area of the universe at
low-redshifts and determined the position of each galaxies. Such map
of galaxies implies the existence of large filamentary structures and
can provide the fruitful information to take aim at the location of
filaments \citep{Erdogdu:2004, Tempel:2014}. At higher redshifts one
can resort in principle to Ly-$\alpha$ forest data.

\subsubsection{Galactic Foregrounds:}  
Diffuse foregrounds at low frequencies are very prominent and might be
orders of magnitude higher than the signal. For example, in the range
of $\approx 150$~MHz the foregrounds -- which are dominated by diffuse
synchrotron radiation and extra galactic point sources confusion --
are about 3 orders of magnitude higher than the \HI\ signal
\citep{Shaver:1999, Jelic:2008, Jelic:2010, Bernardi:2009, Bernardi:2010}.
However, even at such low frequencies the smooth
spectral properties of the Galactic diffuse foregrounds and
extra-galactic point sources allow their disentanglement from the
cosmological signal of interest 
\citep{Jelic:2008, Harker:2009, Chapman:2012, Chapman:2013}.  
At higher frequencies the intensity of the foregrounds drops vey
significantly; typically as $\propto \nu^{-2.6}$ \citep{Shaver:1999}
which reduces the problem of foregrounds significantly around 1~GHz.

In addition to the different frequency dependent behavior of the
signal from filaments with respect to the diffuse and point source
foregrounds, their disentanglement requires a late bandwidth within
which the two behaviors can be clearly seen.
As an example we consider a $\Delta l \propto 100$~cMpc long filament
elongated along the line of sight at $z\approx 1$.  The signal from
this filament has has a frequency range of $ \Delta\nu/\nu \approx
\Delta v/c = H(z) \Delta l /c \approx 7\% $ which translates to less
than 100~MHz.  Such frequency range is much smaller than the
observational bandwidth that current and future instrument have. Such
an argument works over the whole range of redshifts that we have
considered in this paper.
Therefore, the foregrounds are a surmountable obstacle and will not
pose a significant problem to the prospect of observing filaments with
\HI\ and \HeII.

\subsubsection{Contamination from Virialized objects}
Quantifying the contribution of \HI\ and \HeII\ emission from galaxies
or other virtualized is very difficult as there is very little data on
\HI\ beyond $z\approx 0$ and no data whatsoever on \HeII.
\citet{Obreschkow:2009} have simulated the expected source counts on
the sky as a function of redshift where they estimated the number of
radio sources above a certain flux limit as a function of redshift per
squared degree within redshift bins of $\Delta z=0.1$,
${dN/dz}/{\mathrm{deg^2}}$ (see Figure 5 in their paper).  Using this
figure, the number of sources above 1~$\mu$Jy increases as $z^{2.3}$
at low redshifts; peaks at $z=0.5$ and then drops exponentially at at
higher redshifts. The peak value reaches a few time $10^5$ sources
which for a beam of a few arc minutes drops down to $10^3$ at most.
This is of course something that one should worry about, however,
given that these source are very highly localized in frequency 
(\citet{Obreschkow:2009} adopted a value of 8 km/sec width) it is
straightforward to show that with a high frequency resolution data one
can filter these objects by applying a low pass filter along the
frequency direction. We plan to address this point in more detail in
future work.

\section{Conclusions}
\label{sec:conc}

Observing the hyperfine transition of the intergalactic \HI\ and
$^3$\HeII\ provides a useful probe for baryons in the high density
regions of the diffuse IGM. These probes can help shed light on many
problems, such as, the missing baryons and the large-scale structure
on the quasi-linear scales.  In this paper we estimate the prospects
for observing \HI\ and \iHeII\ in the intergalactic filamentary
structures using their respective hyperfine transitions.

The computation of the expected \HI\ and \iHeII\ signal requires a
precise estimation for the ionization state of each component and
physical processes determining their spin temperatures.  These
calculations have been carried out assuming equilibrium radiative
transport equations \citep{Fukugita:1994} and the UV/X-ray background
model of \citet{Haardt:2012}.
Two types of filament models are considered: simple elongated slabs
with varying baryon density, and filaments extracted from
cosmologically simulated large scale filaments.  The estimated signals
of \HI\ and $^3$\HeII\ obtained are generally weak at
low-redshifts. This is because hydrogen and helium reionization
processes have both been completed. The amplitude of the brightness
temperature of \HI\ is $\dTb \sim 10^{-6}-10^{-3}$ K and of
$^3$\HeII\ is $\dTb \sim 10^{-10}-10^{-6}$K at $z=0.5$. Even at $z=1$,
the signals of both \HI\ and $^3$\HeII\ increase only by a factor of a
few.

We have also discussed whether current telescopes can reach the
sensitivities required for observing \HI\ and \iHeII\ emission from
our filament models.  Despite the low signal of \HI, we have shown in
this paper that these levels of sensitivity can be reached even by the
present instruments, such as GBT, EVLA, and Arecibo with 100 hours of
integration time.  The $^3$\HeII\ signal is much weaker than that of
\HI\ and is very hard to detect with present instruments.  However, we
show that detecting both \HI\ and \iHeII\ from large scale filaments
is possible with future telescopes, such as the full MeerKAT and SKA,
given a reasonable amount of integration time.

At higher redshifts one benefits from larger signal, however, other
issues start to become important. Most importantly, the beam size at
higher frequencies becomes smaller, along with the field of view. The
smaller beam size means that more distant filaments will not fill the
beam, leading to a smaller detected signal.  The field of view issue
will limit the size of the filaments we can detect, especially if the
filament is mostly elongated perpendicular to the line-of-sight.
However, as we show the $^3$\HeII\ signal around $z \sim 4$ is still
in principle detectable even with the present instrument such as GMRT
and Arecibo.

Obviously, one of the benefits of observing the two lines in the same
filament is to give us a handle on the UV/X-ray background. In this
work, we have adopted a single model for the background, but in future
work this aspect will be explored in more detail.

In this paper we focus only on the direct detection of these signals.
Less direct detections, e.g., statistical methods or cross-correlation
approaches are also possible and, in some cases of \HI\, been
discussed by other authors \citep{Chang:2008}.  This is clearly an
exciting route that promise to yield results in the foreseeable
future.
One can consider also detection of the signal in absorption towards
background radio sources would be in principle advantageous to the
emission signal at low redshifts where the spin temperatures are low
\citep{McQuinn:2009b}. This is of course depends on the number density
of radio sources at high redshifts against which this measurement can
be in principle done. We defer this to a future work.

It should be mentioned that we have not considered a number of effects
that might make it more difficult to observe the signal, e.g.,
foregrounds, contamination from recombination lines, metal lines, etc.
Fitting out the foregrounds will be necessary and the removing long
wavelength modes along the line of sight is a crucial. This can be
done provided that the bandwidth of the observation is much larger
than the bandwidth covered by the filament. Such a requirement is
easily satisfied with the current and future generations telescopes.
In Addition, the point sources can also contaminate the observation,
however these sources are very localized and can be removed by
filtering them out using high angular resolution data, which is also
available for current and future telescopes.
These two issue will be dealt with in more detail in a future
publication.


\section*{Acknowledgments}
We would like to thank A.~G. de Bruyn for useful discussions and
Elizabeth Fernandez for comments on the manuscript. We acknowledge
support from JSPS (Japan Society for Promotion of Science) fellowship
(YT); JSPS Grant-in-Aid for Scientific Research under Grants
No. 25287057 (NS); Kobayashi-Maskawa Institute for the Origin of
Particles and the Universe; and the World Premier International
Research Center Initiative (WPI Initiative), MEXT, Japan. SZ
acknowledges the Netherlands Organization for Scientific Research
(NWO) VICI grant for financial support. This work was supported by the
JSPS Strategic Young Researcher Overseas Visits Program for
Accelerating Brain Circulation.

\bibliography{astro}

\begin{thebibliography}{}

\bibitem[\protect\citeauthoryear{{Bagla} \& {Loeb}}{{Bagla} \&
  {Loeb}}{2009}]{Bagla:2009}
{Bagla} J.~S.,  {Loeb} A.,  2009, {ArXiv:0905.1698}

\bibitem[\protect\citeauthoryear{{Bell}}{{Bell}}{2000}]{Bell:2000}
{Bell} M.~B.,  2000, \apj, 531, 820

\bibitem[\protect\citeauthoryear{{Bernardi}, {de Bruyn}, {Brentjens}, {Ciardi},
  {Harker}, {Jeli{\'c}}, {Koopmans}, {Labropoulos}, {Offringa}, {Pandey},
  {Schaye}, {Thomas}, {Yatawatta} \& {Zaroubi}}{{Bernardi}
  et~al.}{2009}]{Bernardi:2009}
{Bernardi} G.,  {de Bruyn} A.~G.,  {Brentjens} M.~A.,  {Ciardi} B.,  {Harker}
  G.,  {Jeli{\'c}} V.,  {Koopmans} L.~V.~E.,  {Labropoulos} P.,  {Offringa} A.,
   {Pandey} V.~N.,  {Schaye} J.,  {Thomas} R.~M.,  {Yatawatta} S.,    {Zaroubi}
  S.,  2009, \aap, 500, 965

\bibitem[\protect\citeauthoryear{{Bernardi}, {de Bruyn}, {Harker}, {Brentjens},
  {Ciardi}, {Jeli{\'c}}, {Koopmans}, {Labropoulos}, {Offringa}, {Pandey},
  {Schaye}, {Thomas}, {Yatawatta} \& {Zaroubi}}{{Bernardi}
  et~al.}{2010}]{Bernardi:2010}
{Bernardi} G.,  {de Bruyn} A.~G.,  {Harker} G.,  {Brentjens} M.~A.,  {Ciardi}
  B.,  {Jeli{\'c}} V.,  {Koopmans} L.~V.~E.,  {Labropoulos} P.,  {Offringa} A.,
   {Pandey} V.~N.,  {Schaye} J.,  {Thomas} R.~M.,  {Yatawatta} S.,    {Zaroubi}
  S.,  2010, \aap, 522, A67

\bibitem[\protect\citeauthoryear{{Black}}{{Black}}{1981}]{Black:1981a}
{Black} J.~H.,  1981, \mnras, 197, 553

\bibitem[\protect\citeauthoryear{{Bond}, {Kofman} \& {Pogosyan}}{{Bond}
  et~al.}{1996}]{Bond:1996}
{Bond} J.~R.,  {Kofman} L.,    {Pogosyan} D.,  1996, \nat, 380, 603

\bibitem[\protect\citeauthoryear{Cen}{Cen}{1992}]{Cen:1992c}
Cen R.,  1992, Astrophys.J.Suppl., 78, 341

\bibitem[\protect\citeauthoryear{{Cen} \& {Ostriker}}{{Cen} \&
  {Ostriker}}{1999}]{Cen:1999}
{Cen} R.,  {Ostriker} J.~P.,  1999, \apj, 514, 1

\bibitem[\protect\citeauthoryear{{Chang}, {Pen}, {Bandura} \&
  {Peterson}}{{Chang} et~al.}{2010}]{Chang:2010}
{Chang} T.-C.,  {Pen} U.-L.,  {Bandura} K.,    {Peterson} J.~B.,  2010, ArXiv:1007.3709


\bibitem[\protect\citeauthoryear{{Chang}, {Pen}, {Peterson} \&
  {McDonald}}{{Chang} et~al.}{2008}]{Chang:2008}
{Chang} T.-C.,  {Pen} U.-L.,  {Peterson} J.~B.,    {McDonald} P.,  2008,
  Physical Review Letters, 100, 091303

\bibitem[\protect\citeauthoryear{Chapman, Abdalla, Bobin, Starck, Harker
  et~al.,}{Chapman et~al.}{2012}]{Chapman:2012}
Chapman E.,  Abdalla F.~B.,  Bobin J.,  Starck J.-L.,  Harker G.,    et~al.,
  2012

\bibitem[\protect\citeauthoryear{{Chapman}, {Abdalla}, {Bobin}, {Starck},
  {Harker}, {Jeli{\'c}}, {Labropoulos}, {Zaroubi}, {Brentjens}, {de Bruyn} \&
  {Koopmans}}{{Chapman} et~al.}{2013}]{Chapman:2013}
{Chapman} E.,  {Abdalla} F.~B.,  {Bobin} J.,  {Starck} J.-L.,  {Harker} G.,
  {Jeli{\'c}} V.,  {Labropoulos} P.,  {Zaroubi} S.,  {Brentjens} M.~A.,  {de
  Bruyn} A.~G.,    {Koopmans} L.~V.~E.,  2013, \mnras, 429, 165

\bibitem[\protect\citeauthoryear{{Choi}, {Bond}, {Strauss}, {Coil}, {Davis} \&
  {Willmer}}{{Choi} et~al.}{2010}]{Choi:2010}
{Choi} E.,  {Bond} N.~A.,  {Strauss} M.~A.,  {Coil} A.~L.,  {Davis} M.,
  {Willmer} C.~N.~A.,  2010, \mnras, 406, 320

\bibitem[\protect\citeauthoryear{{Colless}, {Dalton}, {Maddox}, {Sutherland},
  {Norberg}, {Cole}, {Bland-Hawthorn}, {Bridges} \& {Cannon}}{{Colless}
  et~al.}{2001}]{Colless:2001}
{Colless} M.,  {Dalton} G.,  {Maddox} S.,  {Sutherland} W.,  {Norberg} P.,
  {Cole} S.,  {Bland-Hawthorn} J.,  {Bridges} T.,    {Cannon} R.,  2001,
  \mnras, 328, 1039

\bibitem[\protect\citeauthoryear{{Erdo{\~ g}du}, {Lahav}, {Zaroubi},
  {Efstathiou}, {Moody}, {Peacock}, {Colless}, {Baldry}, {Baugh} \&
  {Bland-Hawthorn}}{{Erdo{\~ g}du} et~al.}{2004}]{Erdogdu:2004}
{Erdo{\~ g}du} P.,  {Lahav} O.,  {Zaroubi} S.,  {Efstathiou} G.,  {Moody} S.,
  {Peacock} J.~A.,  {Colless} M.,  {Baldry} I.~K.,  {Baugh} C.~M.,
  {Bland-Hawthorn} J.,  2004, \mnras, 352, 939

\bibitem[\protect\citeauthoryear{{Ewen} \& {Purcell}}{{Ewen} \&
  {Purcell}}{1951}]{Ewen:1951}
{Ewen} H.~I.,  {Purcell} E.~M.,  1951, \nat, 168, 356

\bibitem[\protect\citeauthoryear{{Field}}{{Field}}{1958}]{Field:1958}
{Field} G.~B.,  1958, Proceedings of the IRE, 46, 240

\bibitem[\protect\citeauthoryear{{Field}}{{Field}}{1959}]{Field:1959b}
{Field} G.~B.,  1959, \apj, 129, 536

\bibitem[\protect\citeauthoryear{{Freudling}, {Staveley-Smith}, {Catinella},
  {Minchin}, {Calabretta}, {Momjian}, {Zwaan}, {Meyer} \& {O'Neil}}{{Freudling}
  et~al.}{2011}]{Freudling:2011}
{Freudling} W.,  {Staveley-Smith} L.,  {Catinella} B.,  {Minchin} R.,
  {Calabretta} M.,  {Momjian} E.,  {Zwaan} M.,  {Meyer} M.,    {O'Neil} K.,
  2011, \apj, 727, 40

\bibitem[\protect\citeauthoryear{{Fukugita} \& {Kawasaki}}{{Fukugita} \&
  {Kawasaki}}{1994}]{Fukugita:1994}
{Fukugita} M.,  {Kawasaki} M.,  1994, \mnras, 269, 563

\bibitem[\protect\citeauthoryear{{Fukugita} \& {Peebles}}{{Fukugita} \&
  {Peebles}}{2004}]{Fukugita:2004}
{Fukugita} M.,  {Peebles} P.~J.~E.,  2004, \apj, 616, 643

\bibitem[\protect\citeauthoryear{Furlanetto \& Furlanetto}{Furlanetto \&
  Furlanetto}{007b}]{Furlanetto:2007b}
Furlanetto S.,  Furlanetto M.,  2007b, Mon.Not.Roy.Astron.Soc., 379, 130

\bibitem[\protect\citeauthoryear{Furlanetto, Oh \& Briggs}{Furlanetto
  et~al.}{2006}]{Furlanetto:2006}
Furlanetto S.,  Oh S.~P.,    Briggs F.,  2006, Phys.Rept., 433, 181

\bibitem[\protect\citeauthoryear{{Furlanetto} \& {Furlanetto}}{{Furlanetto} \&
  {Furlanetto}}{007a}]{Furlanetto:2007a}
{Furlanetto} S.~R.,  {Furlanetto} M.~R.,  2007a, \mnras, 374, 547

\bibitem[\protect\citeauthoryear{Furlanetto, Oh \& Pierpaoli}{Furlanetto
  et~al.}{2006}]{Furlanetto:2006c}
Furlanetto S.~R.,  Oh S.~P.,    Pierpaoli E.,  2006, Phys.Rev., D74, 103502

\bibitem[\protect\citeauthoryear{{Furlanetto} \& {Pritchard}}{{Furlanetto} \&
  {Pritchard}}{2006}]{Furlanetto:2006a}
{Furlanetto} S.~R.,  {Pritchard} J.~R.,  2006, \mnras, 372, 1093

\bibitem[\protect\citeauthoryear{{Goldwire} Jr. \& {Goss}}{{Goldwire} \&
  {Goss}}{1967}]{Goldwire:1967}
{Goldwire} Jr. H.~C.,  {Goss} W.~M.,  1967, \apj, 149, 15

\bibitem[\protect\citeauthoryear{{Gould}}{{Gould}}{1994}]{Gould:1994}
{Gould} R.~J.,  1994, \apj, 423, 522

\bibitem[\protect\citeauthoryear{{Haardt} \& {Madau}}{{Haardt} \&
  {Madau}}{2012}]{Haardt:2012}
{Haardt} F.,  {Madau} P.,  2012, \apj, 746, 125

\bibitem[\protect\citeauthoryear{{Harker}, {Zaroubi}, {Bernardi}, {Brentjens},
  {de Bruyn}, {Ciardi}, {Jeli{\'c}}, {Koopmans}, {Labropoulos}, {Mellema},
  {Offringa}, {Pandey}, {Schaye}, {Thomas} \& {Yatawatta}}{{Harker}
  et~al.}{2009}]{Harker:2009}
{Harker} G.,  {Zaroubi} S.,  {Bernardi} G.,  {Brentjens} M.~A.,  {de Bruyn}
  A.~G.,  {Ciardi} B.,  {Jeli{\'c}} V.,  {Koopmans} L.~V.~E.,  {Labropoulos}
  P.,  {Mellema} G.,  {Offringa} A.,  {Pandey} V.~N.,  {Schaye} J.,  {Thomas}
  R.~M.,    {Yatawatta} S.,  2009, \mnras, 397, 1138

\bibitem[\protect\citeauthoryear{{Hirata}}{{Hirata}}{2006}]{Hirata:2006}
{Hirata} C.~M.,  2006, \mnras, 367, 259

\bibitem[\protect\citeauthoryear{{Jeli{\'c}}, {Zaroubi}, {Labropoulos},
  {Bernardi}, {de Bruyn} \& {Koopmans}}{{Jeli{\'c}} et~al.}{2010}]{Jelic:2010}
{Jeli{\'c}} V.,  {Zaroubi} S.,  {Labropoulos} P.,  {Bernardi} G.,  {de Bruyn}
  A.~G.,    {Koopmans} L.~V.~E.,  2010, \mnras, 409, 1647

\bibitem[\protect\citeauthoryear{{Jeli{\'c}}, {Zaroubi}, {Labropoulos},
  {Thomas}, {Bernardi}, {Brentjens}, {de Bruyn}, {Ciardi}, {Harker},
  {Koopmans}, b {Pandey}, {Schaye} \& {Yatawatta}}{{Jeli{\'c}}
  et~al.}{2008}]{Jelic:2008}
{Jeli{\'c}} V.,  {Zaroubi} S.,  {Labropoulos} P.,  {Thomas} R.~M.,  {Bernardi}
  G.,  {Brentjens} M.~A.,  {de Bruyn} A.~G.,  {Ciardi} B.,  {Harker} G.,
  {Koopmans} L.~V.~E.,  b {Pandey} V.~N.,  {Schaye} J.,    {Yatawatta} S.,
  2008, \mnras, 389, 1319

\bibitem[\protect\citeauthoryear{{Kaiser}}{{Kaiser}}{1987}]{Kaiser:1987}
{Kaiser} N.,  1987, \mnras, 227, 1

\bibitem[\protect\citeauthoryear{{Komatsu}, {Smith}, {Dunkley}, {Bennett},
  {Gold}, {Hinshaw}, {Jarosik}, {Larson}, {Nolta}, {Page}, {Spergel},
  {Halpern}, {Hill}, {Kogut} \& {Limon}}{{Komatsu} et~al.}{2011}]{Komatsu:2011}
{Komatsu} E.,  {Smith} K.~M.,  {Dunkley} J.,  {Bennett} C.~L.,  {Gold} B.,
  {Hinshaw} G.,  {Jarosik} N.,  {Larson} D.,  {Nolta} M.~R.,  {Page} L.,
  {Spergel} D.~N.,  {Halpern} M.,  {Hill} R.~S.,  {Kogut} A.,    {Limon} M.,
  2011, \apjs, 192, 18

\bibitem[\protect\citeauthoryear{McQuinn \& Switzer}{McQuinn \&
  Switzer}{2009}]{McQuinn:2009b}
McQuinn M.,  Switzer E.~R.,  2009, Phys.Rev., D80, 063010

\bibitem[\protect\citeauthoryear{{Madau}, {Meiksin} \& {Rees}}{{Madau}
  et~al.}{1997}]{Madau:1997}
{Madau} P.,  {Meiksin} A.,    {Rees} M.~J.,  1997, \apj, 475, 429

\bibitem[\protect\citeauthoryear{Masui, Switzer, Banavar, Bandura, Blake
  et~al.,}{Masui et~al.}{2012}]{Masui:2012}
Masui K.,  Switzer E.,  Banavar N.,  Bandura K.,  Blake C.,    et~al., 2012

\bibitem[\protect\citeauthoryear{Matsuda, Sato \& Takeda}{Matsuda
  et~al.}{1971}]{Matsuda:1971}
Matsuda T.,  Sato H.,    Takeda H.,  1971, Prog.Theor.Phys., 46, 416

\bibitem[\protect\citeauthoryear{{Menzel} \& {Pekeris}}{{Menzel} \&
  {Pekeris}}{1935}]{Menzel:1935}
{Menzel} D.~H.,  {Pekeris} C.~L.,  1935, \mnras, 96, 77

\bibitem[\protect\citeauthoryear{Mo, van~den Bosch \& White}{Mo
  et~al.}{2010}]{MoWhite:2010}
Mo H.,  van~den Bosch F.,    White S.,  2010, Galaxy Formation and Evolution, 1
  edn.
Cambridge University Press

\bibitem[\protect\citeauthoryear{{Muller} \& {Oort}}{{Muller} \&
  {Oort}}{1951}]{Muller:1951}
{Muller} C.~A.,  {Oort} J.~H.,  1951, \nat, 168, 357

\bibitem[\protect\citeauthoryear{{Nan}, {Li}, {Jin}, {Wang}, {Zhu}, {Zhu},
  {Zhang}, {Yue} \& {Qian}}{{Nan} et~al.}{2011}]{Nan:2011}
{Nan} R.,  {Li} D.,  {Jin} C.,  {Wang} Q.,  {Zhu} L.,  {Zhu} W.,  {Zhang} H.,
  {Yue} Y.,    {Qian} L.,  2011, International Journal of Modern Physics D, 20,
  989

\bibitem[\protect\citeauthoryear{{Nussbaumer} \& {Storey}}{{Nussbaumer} \&
  {Storey}}{1983}]{Nussbaumer:1983}
{Nussbaumer} H.,  {Storey} P.~J.,  1983, \aap, 126, 75

\bibitem[\protect\citeauthoryear{{Obreschkow}, {Kl{\"o}ckner}, {Heywood},
  {Levrier} \& {Rawlings}}{{Obreschkow} et~al.}{2009}]{Obreschkow:2009}
{Obreschkow} D.,  {Kl{\"o}ckner} H.-R.,  {Heywood} I.,  {Levrier} F.,
  {Rawlings} S.,  2009, \apj, 703, 1890

\bibitem[\protect\citeauthoryear{{Osterbrock}}{{Osterbrock}}{1989}]{Osterbrock:1989}
{Osterbrock} D.~E.,  1989, {Astrophysics of gaseous nebulae and active galactic
  nuclei}

\bibitem[\protect\citeauthoryear{{Pandey}, {Kulkarni}, {Bharadwaj} \&
  {Souradeep}}{{Pandey} et~al.}{2011}]{Pandey:2011}
{Pandey} B.,  {Kulkarni} G.,  {Bharadwaj} S.,    {Souradeep} T.,  2011, \mnras,
  411, 332

\bibitem[\protect\citeauthoryear{{Popping} \& {Braun}}{{Popping} \&
  {Braun}}{2011}]{Popping:2011}
{Popping} A.,  {Braun} R.,  2011, \aap, 527, A90

\bibitem[\protect\citeauthoryear{{Popping}, {Dav{\'e}}, {Braun} \&
  {Oppenheimer}}{{Popping} et~al.}{2009}]{Popping:2009}
{Popping} A.,  {Dav{\'e}} R.,  {Braun} R.,    {Oppenheimer} B.~D.,  2009, \aap,
  504, 15

\bibitem[\protect\citeauthoryear{{Pritchard} \& {Furlanetto}}{{Pritchard} \&
  {Furlanetto}}{2006}]{Pritchard:2006}
{Pritchard} J.~R.,  {Furlanetto} S.~R.,  2006, \mnras, 367, 1057

\bibitem[\protect\citeauthoryear{{Pritchard} \& {Loeb}}{{Pritchard} \&
  {Loeb}}{2012}]{Pritchard:2012}
{Pritchard} J.~R.,  {Loeb} A.,  2012, Reports on Progress in Physics, 75,
  086901

\bibitem[\protect\citeauthoryear{{Rood}, {Wilson} \& {Steigman}}{{Rood}
  et~al.}{1979}]{Rood:1979}
{Rood} R.~T.,  {Wilson} T.~L.,    {Steigman} G.,  1979, \apjl, 227, L97

\bibitem[\protect\citeauthoryear{{Shaver}, {Windhorst}, {Madau} \& {de
  Bruyn}}{{Shaver} et~al.}{1999}]{Shaver:1999}
{Shaver} P.~A.,  {Windhorst} R.~A.,  {Madau} P.,    {de Bruyn} A.~G.,  1999,
  \aap, 345, 380

\bibitem[\protect\citeauthoryear{Shchekinov \& Vasiliev}{Shchekinov \&
  Vasiliev}{2007}]{Shchekinov:2007}
Shchekinov Y.~A.,  Vasiliev E.,  2007, Mon.Not.Roy.Astron.Soc., 379, 1003

\bibitem[\protect\citeauthoryear{{Sigurdson} \& {Furlanetto}}{{Sigurdson} \&
  {Furlanetto}}{2006}]{Sigurdson:2006}
{Sigurdson} K.,  {Furlanetto} S.~R.,  2006, Physical Review Letters, 97, 091301

\bibitem[\protect\citeauthoryear{{Skrutskie}, {Cutri}, {Stiening}, {Weinberg},
  {Schneider}, {Carpenter}, {Beichman}, {Capps} \& {Chester}}{{Skrutskie}
  et~al.}{2006}]{Skrutskie:2006}
{Skrutskie} M.~F.,  {Cutri} R.~M.,  {Stiening} R.,  {Weinberg} M.~D.,
  {Schneider} S.,  {Carpenter} J.~M.,  {Beichman} C.,  {Capps} R.,    {Chester}
  T.,  2006, \aj, 131, 1163

\bibitem[\protect\citeauthoryear{Spitzer}{Spitzer}{1978}]{Spitzer:1978}
Spitzer L.,  1978, Physical Processes In The Interstellar Medium.
Wiley classics library, Wiley \& Son

\bibitem[\protect\citeauthoryear{{Springel}}{{Springel}}{2005}]{Springel:2005}
{Springel} V.,  2005, \mnras, 364, 1105

\bibitem[\protect\citeauthoryear{{Sunyaev}}{{Sunyaev}}{1966}]{Sunyaev:1966}
{Sunyaev} R.~A.,  1966, \azh, 43, 1237

\bibitem[\protect\citeauthoryear{{Switzer}, {Masui}, {Bandura}, {Calin},
  {Chang}, {Chen}, {Li}, {Liao}, {Natarajan}, {Pen}, {Peterson}, {Shaw} \&
  {Voytek}}{{Switzer} et~al.}{2013}]{Switzer:2013}
{Switzer} E.~R.,  {Masui} K.~W.,  {Bandura} K.,  {Calin} L.-M.,  {Chang} T.-C.,
   {Chen} X.-L.,  {Li} Y.-C.,  {Liao} Y.-W.,  {Natarajan} A.,  {Pen} U.-L.,
  {Peterson} J.~B.,  {Shaw} J.~R.,    {Voytek} T.~C.,  2013, ArXiv:1304.3712

\bibitem[\protect\citeauthoryear{{Tempel}, {Stoica}, {Mart{\'{\i}}nez},
  {Liivam{\"a}gi}, {Castellan} \& {Saar}}{{Tempel} et~al.}{2014}]{Tempel:2014}
{Tempel} E.,  {Stoica} R.~S.,  {Mart{\'{\i}}nez} V.~J.,  {Liivam{\"a}gi} L.~J.,
   {Castellan} G.,    {Saar} E.,  2014, \mnras, 438, 3465

\bibitem[\protect\citeauthoryear{{Townes}}{{Townes}}{1957}]{Townes:1957}
{Townes} C.~H.,  1957, in {van de Hulst} H.~C.,  ed., Radio astronomy Vol.~4 of
  IAU Symposium, {Microwave and radio-frequency resonance lines of interest to
  radio astronomy}.
p.~92

\bibitem[\protect\citeauthoryear{{van de Hulst}}{{van de
  Hulst}}{1945}]{Hulst:1945}
{van de Hulst} H.,  1945, Nederlands Tijdschrift voor Natuuurkunde, 11, 210

\bibitem[\protect\citeauthoryear{{van Haarlem}, {Wise}, {Gunst} \& et al.}{{van
  Haarlem} et~al.}{2013}]{Haarlem:2013}
{van Haarlem} M.~P.,  {Wise} M.~W.,  {Gunst}   et al. 2013, \aap, 556, A2

\bibitem[\protect\citeauthoryear{{Verner} \& {Ferland}}{{Verner} \&
  {Ferland}}{1996}]{Verner:1996}
{Verner} D.~A.,  {Ferland} G.~J.,  1996, \apjs, 103, 467

\bibitem[\protect\citeauthoryear{{Wouthuysen}}{{Wouthuysen}}{1952}]{Wouthuysen:1952}
{Wouthuysen} S.~A.,  1952, \aj, 57, 31

\bibitem[\protect\citeauthoryear{{York}, {Adelman}, {Anderson} Jr. \& {SDSS
  Collaboration}}{{York} et~al.}{2000}]{York:2000}
{York} D.~G.,  {Adelman} J.,  {Anderson} Jr. J.~E.,    {SDSS Collaboration}
  2000, \aj, 120, 1579

\bibitem[\protect\citeauthoryear{{Zaroubi}}{{Zaroubi}}{2013}]{Zaroubi:2013}
{Zaroubi} S.,  2013, in {Wiklind} T.,  {Mobasher} B.,   {Bromm} V.,  eds,
  Astrophysics and Space Science Library Vol.~396 of Astrophysics and Space
  Science Library, {The Epoch of Reionization}.
p.~45

\bibitem[\protect\citeauthoryear{{Zygelman}}{{Zygelman}}{2005}]{Zygelman:2005}
{Zygelman} B.,  2005, \apj, 622, 1356

\end{thebibliography}

\appendix

\section{Recombination and ionization coefficients}
\label{sec:cof}

For the sake of completeness, here we summarize the recombination and
collisional ionization rates and the cooling function adopted in this
work. We take these values from
\citet{Fukugita:1994,MoWhite:2010}. Similar expressions also can be
found in
\citet{Menzel:1935,Matsuda:1971,Spitzer:1978,Black:1981a,Cen:1992c,Verner:1996}.

\subsection{Recombination and Collisional ionization rates}
\label{sec:rates}

\subsection*{Collisional ionization}

\underline{\HI\ $\rightarrow$ \HII } : 
\begin{equation}
\hspace{10pt} 
  \beta_{{\rm H{\small I}}} = 5.85 \times 10^{-11} T^{1/2} 
  \left( 1+T_5^{1/2}\right)^{-1}
  \exp(-1.578 / T_5) 
  \ {\rm cm}^{3} \cdot {\rm sec}^{-1} \, .
\end{equation}
\underline{\HeI\ $\rightarrow$ \HeII } :
\begin{equation}
\hspace{10pt} 
  \beta_{{\rm He{\small I}}} = 2.38 \times 10^{-11} T^{1/2} 
  \left( 1 +  T_5^{1/2} \right)^{-1}
  \exp(-2.853 / T_5) 
  \ {\rm cm}^{3} \cdot {\rm sec}^{-1} \, .
\end{equation}
\underline{\HeII\ $\rightarrow$ \HeIII } :
\begin{equation}
\hspace{10pt} 
  \beta_{{\rm He{\small II}}} = 5.68 \times 10^{-12} T^{1/2} 
  \left( 1 +  T_5^{1/2} \right)^{-1}
  \exp(-6.315 / T_5) 
  \ {\rm cm}^{3} \cdot {\rm sec}^{-1} \, .
\end{equation}

\subsection*{Recombination}

\underline{\HII\ $\rightarrow$ \HI } : (free $\rightarrow$ $n \geq 1$)
\begin{equation}
\hspace{10pt} 
  \alpha_{{\rm H{\small II}}} = 3.96 \times 10^{-13} T_4^{-0.7}
  \left( 1+T_6^{0.7} \right)^{-1} 
  \ {\rm cm}^{3} \cdot {\rm sec}^{-1} \, .
\end{equation}
\underline{\HeII $\rightarrow$ \HeI } : (free $\rightarrow$ $n \geq 1$)
\begin{equation}
\hspace{10pt} 
  \alpha_{{\rm He{\small II}}} = 4.31 \times 10^{-10} T_4^{-0.6353}
  \ {\rm cm}^{3} \cdot {\rm sec}^{-1} \, .
\end{equation}
\underline{\HeIII $\rightarrow$ \HeII } : (free $\rightarrow$ $n \geq 1$)
\begin{equation}
\hspace{10pt} 
  \alpha_{{\rm He{\small III}}} = 2.12 \times 10^{-12} T_4^{-0.7} 
  \left( 1+ 0.379 T_6^{0.7} \right)^{-1} 
  \ {\rm cm}^{3} \cdot {\rm sec}^{-1} \, .
\end{equation}

\subsection*{Dielectric recombination}
\underline{\HeII\ $\rightarrow$ \HeI } :  (dielectric recombination) 
\citep{Nussbaumer:1983,Osterbrock:1989}
\begin{equation}
\hspace{10pt} 
  \xi_{{\rm He{\small II}}} = 6.0 \times 10^{-10} T_5^{-1.5}
  \exp(-4.7 / T_5) 
  \left[ 1 + 0.3 \exp(-0.94 / T_5) \right]
  \ {\rm cm}^{3} \cdot {\rm sec}^{-1} \, .
\end{equation}

\subsection{Cooling function}
\label{sec:cooling}

We define the cooling function $\Lambda$ as;
\begin{equation}
  \Lambda = 
  \sum_{i={\rm HI}, {\rm HeI}, {\rm HeII}} \zeta_i n_e n_i ~~
  + \sum_{i={\rm HI}, {\rm HeI}, {\rm HeII}} \psi_i n_e n_i ~~
  + \sum_{i={\rm HII}, {\rm HeII}, {\rm HeIII}} \eta_i n_e n_i ~~
  + ~~ \omega_{\rm He II} n_e n_{\rm He III} ~~
  + ~~ \lambda_{\rm c} ~~
  + ~~ \theta_{\rm ff}(n_{\rm H II}
  + n_{\rm He II} 
  + 4 n_{\rm He III}) n_e \, , 
\end{equation}
where the right hand side terms are the collisional-ionization
cooling, the collisional excitation cooling, the recombination
cooling, the dielectron recombination cooling, the Compton cooling and
the free-free cooling, respectively.
The symbols $\zeta_i$, $\psi_i$, $\eta_i$, and $\omega_i$ represent
the cooling coefficients due to atomic state $i$, $\lambda_{\rm c}$ is
the Compton cooling rate, and $\theta_{\rm ff}$ is the free-free
cooling coefficient. We summarize the detail expressions of these
values below.

\subsection*{Collisional ionization cooling}
\underline{\HI } :
\begin{equation}
  \hspace{10pt} 
  \zeta_{\rm HI} = 1.27 \times 10^{-21} T^{1/2} \left( 1 + T_5^{1/2} \right)^{-1}
  \exp(-1.58 / T_5) \ \erg \cdot \cm^{3} \cdot \sec^{-1} \, .
\end{equation}
\underline{\HeI } :
\begin{equation}
  \hspace{10pt} 
  \zeta_{\rm HeI} = 9.38 \times 10^{-22} T^{1/2} \left( 1 + T_5^{1/2} \right)^{-1}
  \exp(-2.85 / T_5) \ \erg \cdot \cm^{3} \cdot \sec^{-1} \, .
\end{equation}
\underline{\HeI\ ($2^3$S) } :
\begin{equation}
  \hspace{10pt} 
  \zeta_{\rm HeI, 2^3S} = 5.01 \times 10^{-27} T^{-0.1687} \left( 1 + T_5^{1/2} \right)^{-1}
  \exp(-5.53 / T_4) n_e n_{\rm HeII} / n_{\rm HeI} \ \erg \cdot \cm^{3} \cdot \sec^{-1} \, .
\end{equation}
\underline{\HeII } :
\begin{equation}
  \hspace{10pt} 
  \zeta_{\rm HeII} = 4.95 \times 10^{-22} T^{1/2} \left( 1 + T_5^{1/2} \right)^{-1}
  \exp(-6.31 / T_5) \ \erg \cdot \cm^{3} \cdot \sec^{-1} \, .
\end{equation}

\subsection*{Collisional excitation cooling}
\underline{\HI } :
\begin{equation}
  \hspace{10pt} 
  \psi_{\rm HI} = 7.5 \times 10^{-19} 
  \left( 1 + T_5^{1/2} \right)^{-1}
  \exp(-1.18 / T_5) 
  \ \erg \cdot \cm^{3} \cdot \sec^{-1} \, .
\end{equation}
\underline{\HeI } :
\begin{equation}
  \hspace{10pt} 
  \psi_{\rm HeI} = 9.10 \times 10^{-27} T^{-0.1687} 
  \left( 1 + T_5^{1/2} \right)^{-1}
  \exp(-1.31 / T_4) n_e n_{\rm HeII} / n_{\rm HeI} 
  \ \erg \cdot \cm^{3} \cdot \sec^{-1} \, .
\end{equation}
\underline{\HeII } :
\begin{equation}
  \hspace{10pt} 
  \psi_{\rm HeII} = 5.54 \times 10^{-17} T^{-0.397} 
  \left( 1 + T_5^{1/2} \right)^{-1}
  \exp(-4.73 / T_5) 
  \ \erg \cdot \cm^{3} \cdot \sec^{-1} \, .
\end{equation}

\subsection*{Recombination cooling}
\underline{\HII } :
\begin{equation}
  \hspace{10pt} 
  \eta_{\rm HII} = 2.82 \times 10^{-26} T_3^{0.3} 
  \left( 1 + 3.54 T_6 \right)^{-1}
  \ \erg \cdot \cm^{3} \cdot \sec^{-1} \, .
\end{equation}
\underline{\HeII } :
\begin{equation}
  \hspace{10pt} 
  \eta_{\rm HeII} = 1.55 \times 10^{-26} T^{0.3647} 
  \ \erg \cdot \cm^{3} \cdot \sec^{-1} \, .
\end{equation}
\underline{\HeIII } :
\begin{equation}
  \hspace{10pt} 
  \eta_{\rm HeIII} = 1.49 \times 10^{-25} T^{0.3}
  \left( 1 + 0.885 T_6 \right)^{-1}
  \ \erg \cdot \cm^{3} \cdot \sec^{-1} \, .
\end{equation}

\subsection*{Dielectronic recombination cooling}
\underline{\HeII } :
\begin{equation}
  \hspace{10pt} 
  \omega_{\rm HeII} = 1.24 \times 10^{-13} T^{-1.5} 
  \left( 1 + 0.3 \exp( -9.4/T_4) \right)^{-1}
  \exp(-4.7/T_5) \ \erg \cdot \cm^{3} \cdot \sec^{-1} \, .
\end{equation}

\subsection*{Free-free cooling}
\begin{equation}
  \hspace{10pt} 
  \theta_{\rm ff} = 1.42 \times 10^{-27} g_{\rm ff} T^{1/2} \, ,
\end{equation}
where $g_{\rm ff}$ is the mean Gaunt factor and the values of 
$g_{\rm ff}$ are between 1.1 and 1.5 for $T=10^4$-$10^{8}$ K
\citep{Spitzer:1978} and we adopt the value of $g_{\rm ff}=1.1$ 
in this paper.

\subsection*{Compton cooling}
\begin{equation}
\hspace{10pt}
\lambda_{\rm c} = 4 \kB (T-T_\gamma)\frac{\pi^2}{15}
  \left( \frac{\kB T_\gamma}{\hbar c} \right)^3
  \left( \frac{\kB T_\gamma}{m_e c^2} \right)
  n_e \sigma_{\rm T} c \, .
\end{equation}

\begin{figure}
\begin{center}
\includegraphics[clip,keepaspectratio=true,width=0.45
  \textwidth]{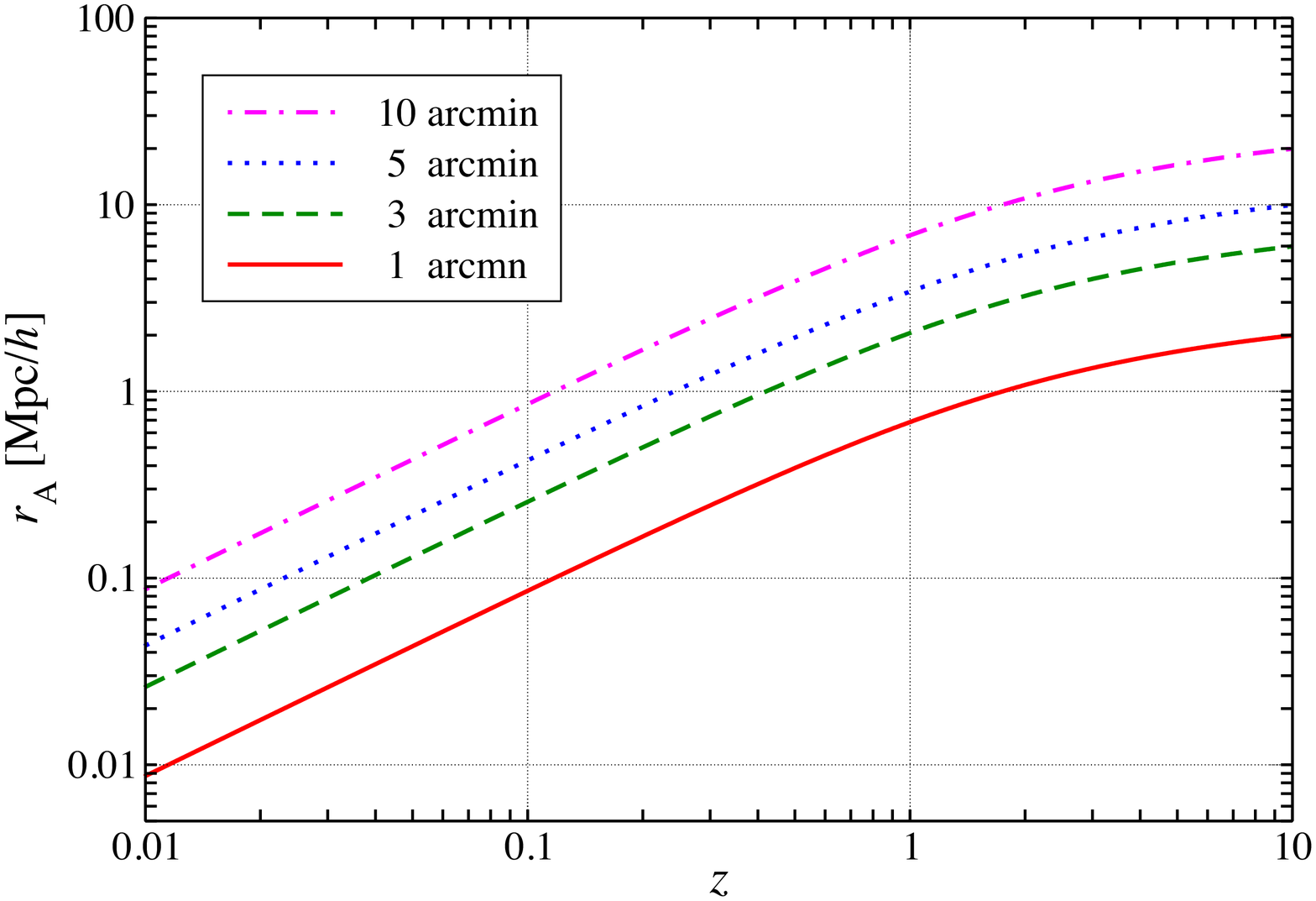}
\hspace{5mm}
\includegraphics[clip,keepaspectratio=true,width=0.45
  \textwidth]{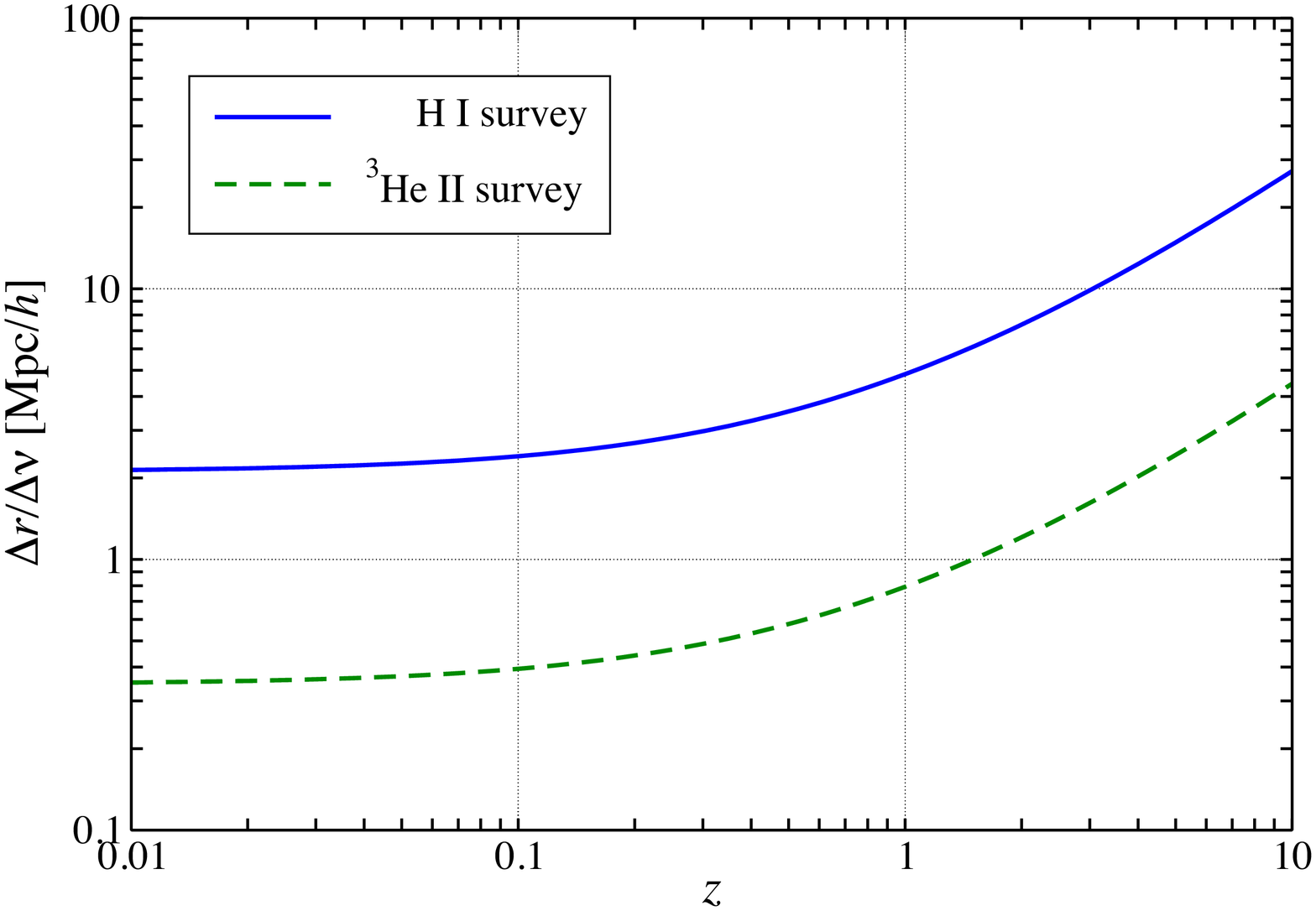}
\caption[]{(Left)The spacial size $r_{\rm A}$ with different values of
  the angular size $\theta_s$ as a function of redshift. (Right)The
  relationship between the frequency band width $\Delta \nu$ and the
  spacial length along the line-of-sight $\Delta r$ for the
  \HI\ (solid-line) and the $^3$\HeII\ (dashed-line) observations,
  respectively. }
\label{fig:angle}
\end{center}
\end{figure}

\section{The comoving distance for a given bandwidth}
\label{sec:size}

The comoving spatial size $r_{\rm A}$ and the depth along the
line-of-sight ${\Delta r}$ for an observation depend on the
cosmological model. We here summarize them and their relationship with
the survey parameters.

When the angular resolution $\theta_s$ is given, the spatial size can be determined by  
\begin{equation}
  r_{\rm A} = \theta_s \chi(z) \, ,
\end{equation} 
where $\chi(z)$ is the comoving distance to the observed source at
redshift $z$.  On the other hand, the depth along the line-of-sight is
determined by the frequency band width $\Delta \nu$ and the
relationship between them is given by
\begin{equation}
  \frac{\Delta r}{\Delta \nu} \simeq 
  \frac{1+z}{\nu_0} \frac{c}{H(a)} \, ,
\end{equation}
where $\Delta \nu$ is the frequency bandwidth and $\nu_0$ is the
rest-frame frequency of the observed source. Therefore this
relationship differs between the \HI\ and $^3$\HeII\ observations.

We show the results with our cosmological model as a function of
redshift in Figure~\ref{fig:angle}. 

\bsp

\label{lastpage}

\end{document}